\documentclass{article}     %  [twocolumn]
\usepackage[dvips]{graphicx}

\usepackage{array}
\usepackage{amsmath}\usepackage{amsfonts}\usepackage{amssymb}
\usepackage{pstricks}\usepackage{pst-plot}
\usepackage{float}%%->set figure placement
\usepackage{mathrsfs}%%->script font
\usepackage{natbib}

\newlength{\intwidth}

\def\eps{\varepsilon}

\def\a{\alpha}\def\b{\beta}\def\g{\gamma}\def\d{\delta}
\def\lam{\lambda}
\def\sig{\sigma}\def\om{\omega}
\def\ups{\upsilon}

\def\D{\Delta}

\def\Lop{ \boldsymbol{\cal{L}} }

\def\Cop{ \boldsymbol{\cal{C}} }

\def\Nop{ \boldsymbol{\cal{N}} }

\def\prob{\mathbb{P}}
\def\bfD{{\bf D}}
\def\bfI{{\bf I}}

\def\bs{\boldsymbol}

\def\pa{\partial}\def\ar{\rightarrow}

\def\rmd{{\rm d}}\def\rme{{\rm e}}

\def\max{{\rm max}}
\def\min{{\rm min}}

\def\miz{\textrm{MIZ}}

\begin{document}

%% ------------------------------------------------------------------------ %%
%
%  TITLE
%
%% ------------------------------------------------------------------------ %%

\title{Towards the inclusion of wave-ice interactions in large-scale models for the Marginal Ice Zone}
%
% e.g., \title{Terrestrial ring current:
% Origin, formation, and decay $\alpha\beta\Gamma\Delta$}
% You may use \\ to break the title over several lines.

%% ------------------------------------------------------------------------ %%
%
%  AUTHORS AND AFFILIATIONS
%
%% ------------------------------------------------------------------------ %%
\author{\large
T. D. Williams, $^{1}$
L. G. Bennetts, $^{2}$ V. A. Squire, $^{3}$
D. Dumont,$^{4}$ and L. Bertino$^{1,5}$}

%\vspace{4pt}

\date{February 2012}

\maketitle

%Use \author{\altaffilmark{}} and \altaffiltext{}

% \altaffilmark will produce footnote;
% matching altaffiltext will appear at bottom of page.
% May use \\ to start a new line.
\begin{center}
$^{1}${Nansen Environmental and Remote Sensing Center, Bergen, Norway.}

$^{2}${School of Mathematical Sciences,
The University of Adelaide, Adelaide, Australia.}

$^{3}${Department of Mathematics and Statistics, University of Otago, Dunedin, New Zealand.}

$^{4}${Universit\'e du Qu\'ebec \`a Rimouski, Institut des sciences de la mer, Rimouski, Canada.}

$^{5}${Bjerknes Center for Climate Research, Bergen, Norway.}
\end{center}

%% ------------------------------------------------------------------------ %%
%
%  ABSTRACT
%
%% ------------------------------------------------------------------------ %%

% >> Do NOT include any \begin...\end commands within
% >> the body of the abstract.

\begin{abstract}
A wave-ice interaction model for the marginal ice zone (MIZ) is reported, which involves both the attenuation of ocean surface waves by sea ice and the concomitant breaking of the ice by waves. It is specifically designed to embed wave-ice interactions in an operational ice/ocean model for the first time. We investigate different methods of including the wave forcing, and different criteria for determining if they cause floes to break. We also investigate and discuss the effects of using various attenuation models, finding that predicted MIZ widths are quite sensitive to the choice of model. Additional sensitivity tests are performed on: (i)~different parameterizations of the floe size distribution (FSD), including the initial FSD used; (ii)~the properties of the wave field; and (iii)~the sea ice properties such as concentration, thickness and breaking strain. Results are relatively insensitive to FSD parameterization but vary noticeably and systematically with its initial configuration, as they do with properties (ii--iii). An additional, somewhat surprising sensitivity, is the degree of influence of the numerical scheme that performs wave attenuation and advection. This is because a na\"{i}ve implementation of spatial and temporal discretizations can cause the waves to be over-attenuated, leading to a reduction of the predicted MIZ width by a substantial factor. Example simulations intended to represent conditions in the Fram Strait in 2007, which exploit reanalyzed wave and ice model data, are shown to conclude the results section. These compare favorably to estimates of MIZ width using concentrations obtained from remotely-sensed passive microwave images.
\end{abstract}

%% ------------------------------------------------------------------------ %%
%
%  BEGIN ARTICLE
%
%% ------------------------------------------------------------------------ %%

% The body of the article must start with a \begin{article} command
%
% \end{article} must follow the references section, before the figures
%  and tables.

%% ------------------------------------------------------------------------ %%
%
%  TEXT
%
%% ------------------------------------------------------------------------ %%

\section{Introduction}

In this paper we present a parameterization of wave-ice interactions for an improved representation of marginal ice zone (MIZ) dynamics. The work is part of a project aimed at enhancing forecasting abilities by creating the first wave-ice component for use in an operational ice/ocean model. The demand for predictions of wave-ice interactions in forecasting models is becoming greater as accessibility to the ice-covered regions of the ocean increases due to the impact of climate change \citep{stephenson-etal2011-access}. Indeed, \citet{prinsenberg-peterson2011-beaufort-breakup} recorded flexural failure induced by swell propagating within multiyear pack ice during the summer of 2009 at very large distances from the ice edge in the Beaufort Sea. While the local sea ice there qualified as being heavily decayed by melting \citep{barber-etal2009-pack-ice}, and thus more fragile, these observations suggest that such events could occur more frequently well within the ice pack in a warmer Arctic. The coupling will be embedded in a HYbrid Coordinate Ocean Model (HYCOM), initially in the Fram Strait but intended for future implementation in more general ice/ocean models and oceanic general circulation models. The main effects that we include in our model are the attenuation of waves as they travel from the open ocean into the MIZ,
and the subsequent strain-induced fracture of the ice caused by the ice-coupled waves.

While the notion and importance of integrating wave-ice interactions into an ice/ocean model was broached more than two decades ago the third author (VAS), it is only recently that a prototype analysis has been published \citep{dumont_etal2011}. That work provides the platform for this investigation. The model used by \citet{dumont_etal2011} is one-dimensional, i.e.\ a transect of the ocean, yet integration into an operational model clearly requires an extension to two-dimensions, i.e.\ the ocean surface. A few different themes have been identified for investigation before this geometrical restriction is tackled. The first theme is the manner in which the wave forcing is included and, in particular, the assumption that there is no interaction between waves with differing frequencies. The second is the floe-breaking criterion employed, and the third is the sensitivity of the model results to the choice of attenuation model. We also perform a sensitivity study on the effects of different parameterizations of the floe size distribution (FSD), of properties of the wave field, and of sea ice properties such as concentration, thickness and breaking strain. These parameters can cause significant, although systematic, variation in the model results, which highlights the importance of having accurate estimates for the values to use, and also the need for more experimental observations to be carried out.

The general study concludes with some numerical considerations like how the choice of the advection-attenuation scheme (AAS), the grid size and the time step used, affect our predictions, before we present some more realistic simulations in the Fram Strait using wave and ice model data from 2007. The MIZ widths predicted by these simulations compare quite favorably to estimates made from concentrations obtained from remotely-sensed passive microwave images (AMSR-E, University of Bremen).

The energy carried by a traveling wave is distributed over a spectrum of frequencies, and it is consistent to interpret the effect of the wave forcing on the ice by integrating over the full spectrum. This approach allows for constructive (and destructive) interference between frequencies, and we shall call it the integrated spectrum method (ISM). The ISM was used in ice-breaking contexts by \citet{langhorne_etal01} and \citet{vaughan-squire2011-fracprob}, and also by \citet{rottier1992-jgr-floepairs} who used it to estimate rates of floe collisions and the resulting (acoustic) noise production.

\citet{langhorne_etal01} estimated the lifetimes of ice-sheets in the presence of waves, using the model of \citet{fox_squire94} to define a relationship between significant wave and strain amplitudes in the ice. This is similar to the approach taken in our study, although we prefer to consider a distribution of strains rather than a single representative strain value. \citet{vaughan-squire2011-fracprob}, on the other hand, describe a method of estimating the width of the MIZ  according to the attenuation coefficient of \citet{squire-etal2009-grl} combined with a strain threshold for ice fracture. This work also has similarities to the current study, particularly in regard to the use of a parametric probability threshold. However, their model does not output the FSD, and consequently is not detailed enough to include in an ocean model.

\citet{dumont_etal2011} did not consider interactions between wave frequencies, based
partly on the resulting algebraic simplifications and partly on the assertion that different wave groups would separate due to dispersive effects (as the group velocity, i.e.\ the speed that the energy propagates at, depends on frequency). We will refer to this approach as the wave group method (WGM).

The computational loads required for the ISM and the WGM are similar. We find in our results section that their model predictions are also quite similar when only breaking resulting from strain failure is allowed for. (Stress induced failure was also considered in the WGM setting by \citet{dumont_etal2011}; see below.) Moreover, when our waves-in-ice model (WIM) is extended to two-dimensions by considering a full directional spectrum, we find there is little difference in complexity between using ISM or WGM. However, because the ISM takes a more orthodox theoretical approach than the WGM, 
and also because it produces a smoother FSD,
we advocate its use in future simulations.

Having advected our waves into the ice and allowed for some attenuation to occur as they travel in from the ice edge, the predicted distribution of wave energy must be mapped to a fracture probability using floe-breaking criteria. \citet{dumont_etal2011} propose two floe-breaking criteria, one based on stress failure and one based on strain failure. The former approximates the horizontal stress in the ice, assuming rigid ice and a monochromatic wave surface, and was intended to allow for the effects of cavitation and wetting. During cavitation, the ice floe does not follow the wave profile exactly and potentially causes a strong localized stress on the floe. The criterion predicts greater stress when the waves are longer, which is unphysical in this regime as ice is relatively unaffected by long waves because of their low slope/curvature, normally small amplitude, and the low velocities they force surface objects to move at. As long waves experience the least attenuation from the presence of ice cover, the stress criterion results in an unphysically wide MIZ. Consequently, in due course we will eliminate the stress criterion from the parameterization. However, in the future it is worth investigating a different method of estimating the likelihood of events like cavitation and wetting occurring, as we find that in general only allowing for strain failure causes an under-prediction of MIZ widths. Nonetheless such events are more likely to be caused by higher-frequency, shorter waves, so any model including them will need to reflect this.

One difficulty with using strain failure to determine the probability of floe-breaking is that
experimental measurements for strain thresholds in sea ice are scarce, whereas
there exists an extensive database on its flexural strength \citep{timco-weeks2010-seaiceprops}. We therefore present a method for converting flexural strength to strain thresholds.
To do this, we must also estimate associated ice mechanical properties such as the Young's modulus and Poisson's ratio, which have been measured less often --- especially at the strain rates relevant to this paper. The final result is a strain threshold parameterized by the brine volume fraction, which is a standard parameter in operational sea ice models.

The primary effect of the ice-cover on a traveling wave is to reduce its intensity exponentially with distance. A relationship between this phenomenon and wave scattering by inhomogeneities in the ice cover, especially at floe edges, has been established. The first model of wave attenuation in the ice-covered ocean was produced by \citet{wadhams1973-jgr-attenswell}. Recent appreciation of the role of multiple scattering and wave coherence in the attenuation phenomenon has generated a number of more sophisticated models \citep{kohout_meylan08,squire-etal2009-grl,bennetts-squire2011-atten}. The result is that there is now a selection of attenuation models available for implementation in one-dimensional wave-ice interaction theory, which can provide an attenuation rate as a function of the properties of the incident wave field and ice cover. In this paper we compare the MIZ-width predictions produced by three different attenuation models and focus on the FSDs they are generated from. The first model is calculated by ensemble averaging using a realistic power law distribution for the FSD. However, this process produces artificial resonances that allow a window of high-frequency (and thus high-curvature) waves to be transmitted with negligible attenuation through the MIZ. The result is that the MIZ becomes far too wide.
The second model uses the method of \citet{bennetts-squire2011-atten},
which calculates the average over all possible wave phases throughout the domain.
This model produces a more realistic level of attenuation but, compared to the most thorough attenuation transect conducted to date (collected by \citet{squire_moore80} in the Bering Sea on compliant sea ice floes less than 0.5\,m thick), it still underestimates the attenuation of long waves. Therefore the third model adds some additional damping empirically in order to attenuate these long waves in accord with how they are observed to diminish.
Both the second and third models predict more realistic MIZ widths than the first, but there is still a factor of 2 difference between them. This highlights the importance of resolving the problem of how long waves should be attenuated theoretically, and also of conducting more experiments to confirm and extend the observations of  \citet{squire_moore80} to different thickness ranges.

\section{Model components}

The interactions between ocean waves and sea ice that we are interested in occur on small to medium scales, but they have a profound effect on the large-scale dynamics and thermodynamics of the sea ice. On a large scale the ice cover deforms in response to stresses imposed by winds and currents, and thin ice categories ($<300$ to 500\,mm) are rapidly embedded into thicker ice through convergence, ice formation and ridging. It is customary to model the ice cover as a uniform viscous-plastic material, with all complexities subsumed into the internal stresses of the associated rheology \citep[e.g.][]{hibler-1979-seaice-model-VP,hunke-dukowicz1997-evp}. The viscous-plastic description and most historical ice model developments are based on observations of the central pack where the ice is compact. It is not representative of the part of the ice cover in contact with the open ocean (the MIZ), in which the presence of waves produces smaller floes and lower surface concentrations. The ice in the MIZ also provides less resistance to external forcing and air-ocean heat fluxes and is modulated by the longer endurance of thin ice types. Separate rheologies for the MIZ have been proposed \citep{shen-etal1986,feltham2005-miz-PTRSA}, as well as parameterizations for floe size-dependent thermodynamical processes \citep{steele-etal1989-air-ice-sea,steele1992-melting}. The HYCOM model, in which the wave-ice component will be embedded, uses wave-ice interactions to delineate the MIZ from the inner ice pack. The collisional rheology of \citet{shen_etal87}, in which the collisions are implicitly driven by wave action, is applied in the MIZ and a standard viscous-plastic rheology is used for the remainder of the ice cover.

The bridge between the two scales is the floe size distribution (FSD), which is parametrized according to the observations of \citet{toyota-etal2006-fsd,toyota-etal2011-floedist}, as discussed in \S\ref{sec-fsd}. Although the viscous-plastic rheology is independent of the floe sizes, the FSD plays a key role at the large scale in determining which rheology to use at a particular point. Correspondingly, the MIZ rheology, has an explicit dependence on the FSD. 

At the medium scale, waves are advected from the ocean and are attenuated by the ice. The waves also induce stresses and strains in the ice that may or may not cause them to break. Breakage alters the FSD, lowering the mean floe length and increasing the amount of attenuation, which is predominantly determined by small scale scattering events at the edges of individual floes. An important aspect of the (numerical) breaking process is the means by which we decide whether breakage will occur at a particular point in space, and this is discussed in \S\ref{sec-WIMs}. It is also strongly influenced by the sea ice properties we use --- notably, the Young's modulus, Poisson's ratio, and flexural strength of the ice, which generally depend on the brine volume fraction (see \S\ref{sec-iceprops}) --- and potentially the way we deal with the issue of fatigue (see \S\ref{sec-fat}). The attenuation models we use are discussed in \S\ref{sec-atten}.

\subsection{Floe size distribution}\label{sec-fsd}

In this section we review recent progress towards understanding the FSD in the ice-covered oceans, and describe how this is used in the wave-ice component of the model. Details of the probability density functions (PDFs) considered in the investigation are given in Appendix \ref{app-pdfs}. Prior to 2006, numerous researchers \cite[e.g.][]{weeks-etal1980-fsd,rothrock_thorndike84,matsushita-etal1985-fsd,holt_martin01,toyota-enamoto2002-fsd} had made observations of floe sizes in Arctic areas. It was found that the FSD generally obeyed a power law (Pareto) distribution, having a PDF given by (\ref{pdf-power-law}) with maximum floe size $D_{1} \to \infty$. However, many of the observations were theoretically problematic as the fitted exponent $\g$ was often greater than 2, which implies an infinite ice area if small floes obey this law \citep{rothrock_thorndike84}. This can also be seen by letting $D_{0} \to 0$ in (\ref{pdf-power-law}),  in equation (\ref{meanA-pl}). However, we note that this can also be avoided if the power law is not used for very small floes ({i.e.\ }by keeping $D_{0}$ non-zero). By a similar argument, $\g$ should be greater than 2 for larger floes (let $D_{1}\ar\infty$ in equation~\ref{meanA-pl} to see this). Furthermore, if $D_{0}=0$ the average floe diameter is also infinite if $\g>1$, and the probability of obtaining a floe size greater than zero (which should be 1) is infinite if $\g>0$. 

To address these problems, \citet{toyota-etal2006-fsd}, using data obtained from the southern Sea of Okhotsk, investigated the FSD for
floes ranging between 1\,m and 1.5\,km in diameter. They found that there was a decrease in the value of $\g$ (to about 1.15) when the floe size dropped below about 40\,m. The regime shift was also observed in Antarctica in the late winter of 2006 and 2007 by \citet{toyota-etal2011-floedist}, based on observations in the northwestern Weddell Sea and off Wilkes Land (around 64$^{\circ}$S, 117$^{\circ}$E) with a helicopter-borne digital video camera. Concurrent ice thickness measurements were also made, using a helicopter-borne electromagnetic (EM) sensor above the Weddell Sea and a video system off Wilkes land.
The regime shift was consistent with the value
\begin{equation}\label{Dcrit}
D_{c}=\left(\frac{\pi^{4}Yh^{3}}{48\rho_\textrm{water}g(1-\nu^{2}) }\right)^{1/4},
\end{equation}
which corresponds to the length below which flexural failure cannot occur \citep{mellor1986-icemech}. (The parameters $Y$, $h$, $\nu$, $\rho_\textrm{water}$ and $g$ are the Young's modulus, thickness and Poisson's ratio for sea ice, the density of the water, and the acceleration due to gravity.)
\citet{toyota-etal2011-floedist} also developed a fractal model that would produce the distribution (\ref{pdf-power-law}) for smaller floes, based on a fragility parameter
\begin{equation}
\Pi=\xi^{\g-2},\ \ \g=2+\log_{\xi}\Pi,
\end{equation}
which is the probability that a floe will break into $\xi^2$ pieces. Note that this model predicts $\g<2$.  A value of $\g>2$ corresponds to $\Pi>1$, which implies that there must be additional breaking mechanisms other than the simple breaking scheme used to derive the small floe distribution. Since the regime shift occurs roughly around diameter $D_{c}$, a possible reason is that the smaller floes are mainly produced by breaking due to flexural waves, with the exception of very small bits of rubble and brash ice formed from
floe-floe collisions. Larger floes are primarily influenced by other processes such as the stresses induced by winds and currents. In addition, large floes may also be the result of smaller floes recombining in periods of low temperature, and low wind and wave activity.

A method for using wave-ice interactions to calculate the FSD was proposed by
\citet{dumont_etal2011}.
They calculated the maximum floe size possible in a given wave field, $D_{\max}$, and assumed that in each grid cell the floe lengths obeyed a single-regime power law (\ref{pdf-power-law}), setting $D_{1}=D_{\max}$ and taking $D_{0}$ to be a set minimum floe length $D_{\min}$.
The resulting distribution was coupled to the fractal
model of
\citet{toyota-etal2011-floedist} in their floe breaking simulations, but only once the value of $D_{\max}$ had been determined.
%, the maximum floe length possible in a given wave \vern{field.
Finally, the expected value of $D$, $\langle D\rangle$, was obtained using a novel formula derived as part of the investigation. An alternative means of calculating $\langle D\rangle$ would have been to use equation~(\ref{meanD-pl}). In their simulations, they used the values $D_{\min}=20$\,m, $\xi=2$, and $\Pi=0.9$.

Here, we choose to modify this approach, by using the observed regime change between different power law exponents that occurs around $D_{c}$. To do this we must modify the coupling between the wave-ice interactions and the FSD. If the wave field is large enough that breaking occurs, $D_{\max}$ reduces, so we increase the probability of having small floes, i.e.\ we increase the parameter $\prob_{0}=\prob(D<D_{c})$ in (\ref{pdf-split-power-law}) using the values for $\g_{1}$, $\g_{2}$, $D_{0}$ and $D_{1}$ below, so that $\prob(D<D_{\max})=\prob_{\max}$ is a high probability that we set to 0.95. ($\prob_{0}$ can be found  analytically.) Thus $D_{\max}$ is no longer an absolute maximum but is now a diameter that most of the floe diameters are smaller than. We then calculate the expected floe diameter $\langle D\rangle$ from (\ref{pdf-split-power-law}), using $D_{0}=D_{\min}=20$\,m, $D_{1}=D_{c}$, and $\g_{1}=1.15$ (which corresponds to fragility $\Pi=0.55$), since this was observed in both the Sea of Okhutsk \citep{toyota-etal2006-fsd} and in Antarctica \citep{toyota-etal2011-floedist}. For the larger floes, we take the exponent $\g_{2}$ to be between $2.1$ and $3$. We do not rule out the possibility that breaking produces $D_{\max}<D_{c}$, but once this happens we do not allow $D_{\max}$ to decrease any further. Furthermore, $D_{\max}$ can never decrease below $D_{\min}$. If $D_{\max}<D_{c}$, it is possible that we need $\prob_{0}=1$ in order that $\prob(D<D_{\max})=\prob_{\max}$. If this happens, we must then reduce $D_{1}$ in (\ref{pdf-split-power-law}) from $D_{c}$, and $D_{1}$ is then an absolute maximum. An additional complication that may occur is that $D_{\max}$ is so large that there are no small floes and $\prob_{0}=0$. In that case we would need to increase $D_{1}$ from $D_{c}$, but instead we assume that $D_{\max}>D_{\rm unif}$,
where $D_{\rm unif}$ is chosen to be a sufficiently large diameter; 200\,m is used here. In this case all the floes in the grid cell are taken to be of length $D_{\rm max}$. We shall refer to $D_{\rm unif}$ as the FSD cutoff length.

Another alternative method of changing the FSD in response to breaking is to make $\prob_{0}$ dependent upon the other parameters in the distribution (\ref{pdf-split-power-law}), e.g.\ by requiring that $p_{S}$ be continuous at $D=D_{c}$ (which it seems to be in the plots shown by \citet{toyota-etal2011-floedist}), and increase the fragility $\Pi$, as \citet{toyota-etal2011-floedist} noted that this increased closer to the ice edge. This might make an interesting investigation for the future, but as it is slightly more complicated than the previous methods because it cannot be done analytically, we do not attempt to do this here.

It was hoped that by using the observed distribution (\ref{pdf-split-power-law}), we would implicitly account for some effects other than breaking that are present in the MIZ, such as refreezing. However, it turns out that the MIZ widths predicted are very insensitive to changing between FSD (\ref{pdf-split-power-law}) and the one used by \citet{dumont_etal2011}. Having said this, some sensitivity to the initial value of $D_{\max}$ used does exist. This is tested in \S\ref{sec-testsens} and \S\ref{sec-real}.

\subsection{Ice properties}\label{sec-iceprops}

\citet{timco-weeks2010-seaiceprops} provide an excellent up-to-date synopsis of our current knowledge about the engineering properties of sea ice. This includes information on flexural strength, Young's modulus and Poisson's ratio, which will be used for the present investigation. We note that there are many reported measurements for the flexural strength and Young's modulus of first-year ice, but far fewer for multi-year ice. Only a very small data set is available for the wave-induced failure strain of sea ice {\it in situ}. The data that do exist suggest a value of approximately $(5\pm 3)\times 10^{-5}$. In contrast, many flexural strength tests have been made due to its importance and use in ice engineering problems. Although we recognize that ice fracture is considerably more complicated, we will utilize the experimental findings on  flexural strength to parameterize the probability of ice breaking due to the {\it strains} caused by wave loading. Sea ice is a heterogeneous aggregate of solid ice, brine, air and solid salts, and thus flexural strength is not a basic material property, but an indicial strength, as non-uniform stress fields are created under flexural loads that require assumptions about material behavior.

\begin{figure}[ht]
%\psfrag{vb, ppt}{$v_{b}$, ppt}
%\psfrag{sc, MPa}{$\sig_{c}$, MPa}
%\psfrag{Y, GPa}{$Y^{*}$, GPa}
%\psfrag{ec}{$\eps_{c}$}
%\includegraphics[width=78mm,height=53mm]{iceprops_v2crop.eps}
\begin{center}
\hspace{-28mm}
\includegraphics[width=108mm]{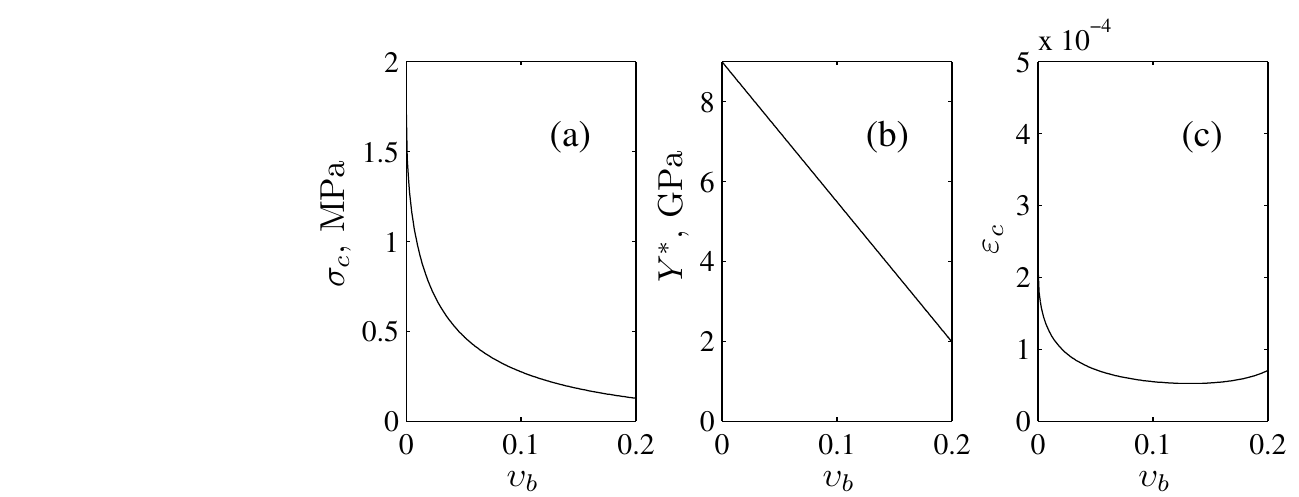}
\caption{Behavior of the flexural strength (a), and our models for the effective Young's modulus (b) and the breaking strain (c) with the brine volume fraction $\upsilon_{b}$.
%{\bf %/home/nersc/timill/validation/matlab/PAPERS\_progs/NERSC/floe-breaking/miz1d:
%iceprops.eps; fig\_iceprops.m}
}
\label{fig-iceprops}
\end{center}
\end{figure}

\citet{timco-obrien1994-flex-strength-fit} collate and analyse 939 flexural strength measurements conducted by 14 different investigators under a variety of conditions and test types, namely, {\it in situ} cantilever tests and simple beam tests with 3- or 4-point loading, to show that the flexural strength $\sigma_c$  has a very simple dependence on  brine volume fraction $\ups_b$:
\begin{equation}\label{sig-c-formula}
\sigma_c = \sigma_0 \exp \left( -5.88 \sqrt{\ups_b} \right),
\end{equation}
where $\sigma_0 = 1.76\rm\,MPa$. This is plotted in Figure~\ref{fig-iceprops}(a), and shows a monotonic decrease from $\sig_{0}$ as $\ups_{b}$ increases. Brine volume is often a parameter in ice-ocean models, but it can also be calculated from the formula of \citet{frankenstein-garner1967-brine}:
\begin{equation}\label{brine-formula}
\ups_b=10^{-3}\frac{s_{\rm ice}}{s_{\rm ice}^{0}}\left(0.532+49.185\times \frac{T_{\rm ice}^{0}}{|T_{\rm ice}|}\right),
\end{equation}
where $s_{\rm ice}$ is the salinity (psu) of the ice and $T_{\rm ice}$ is its temperature ($^{\circ}$C), while $s_{\rm ice}^{0}=1$\,psu and $T_{\rm ice}^{0}=1^{\circ}$C are the unit values of each.

Flexural strength tests are normally analyzed by means of Euler-Bernoulli beam theory, in which the stress normal to the beam cross section, i.e.\ along the beam axis, $\sigma_{11}$, is related to the strain, i.e.\ beam relative extension, $\eps_{11}$, by the one-dimensional Hooke's law $\sigma_{11} = Y\eps_{11}$, where $Y$ is the Young's modulus.
%(See \citet{meirovitch2010-EBref} and Appendix A.) Accordingly, in principle,
Further details are given in Appendix A and by \citet{meirovitch2010-EBref}. In principle, therefore, to convert flexural strength into a breaking strain $\eps_c$ for a beam of sea ice, all we need is $Y$ for sea ice, as $\eps_c=\sigma_c/Y$. However, an ice floe is generally too wide to be modeled as a beam, and is usually modeled as a thin elastic plate, so we must also modify the breaking strain by a factor involving Poisson's ratio $\nu$ (also see Appendix A): $\eps_{c}={\sigma_c}/({Y(1-\nu^{2})}).$

%\vspace{-6mm}

Unfortunately, as \citet{timco-weeks2010-seaiceprops} point out, the constitutive relation for ice is rate-dependent and includes a delayed elastic (anelastic) response (primary, recoverable creep), $\eps^d$, viscous strain (secondary creep) and strain due to cracking (tertiary creep), as well as instantaneous elasticity, $\eps^i$. For modest strains and timescales, $\eps^i$ and $\eps^d$ are of particular importance, as both are impermanent although recovery from delayed elasticity is not instantaneous. In the course of a typical flexural strength test and during the recurring cyclic flexure imparted by ocean surface gravity waves, it is expected that the sea ice will experience stress levels and rates such that the total recoverable strain $\eps^T\approx\eps^i+\eps^d$. This suggests a variation on the instantaneous elastic Young's modulus $Y$ that allows for delayed elasticity to act, that is often called the effective modulus or the strain modulus and that we shall denote by $Y^{*}$. The Young's modulus itself, $Y$, can only be measured dynamically, e.g.\ ultrasonically with small isolated samples or in situ by measuring the propagation of high frequency elastic waves, while $Y^*<Y$ arises from assuming an elastic constitutive relation in an experiment where some recoverable primary creep occurs. Moreover, as with flexural strength, brine volume affects both $Y$ and $Y^*$. \citet{timco-weeks2010-seaiceprops} report a linear relationship for $Y(\ups_b)$ of the form $Y=Y_0(1-3.51\ups_b)$, where $Y_0\approx 10\rm\,GPa$ is roughly the value for freshwater ice at high loading rates. But, whilst increased brine volume leads to a reduction in the effective modulus $Y^*$, the data are too scattered for an empirical relationship for $Y^*(\ups_b)$ to be expressed. For ``average'' brine volumes ranging from 50 to 100\,ppt \cite[$\ups_{b}=0.05$ to 0.1,][]{frankenstein-garner1967-brine}, this suggests $Y$ will reduce to between $\sim 6$--8\,GPa.

As we have noted above, the effect of brine volume on $Y^*$ is more difficult to pin down, but we believe the same kind of reduction would not be unreasonable. More challenging is determining the effect of anelasticity (delayed elasticity) on reducing $Y$ to $Y^*$. The mechanisms that achieve this power-law primary creep with no microcracking cause relaxation processes to occur during cyclical loading, so the rate of loading is important. Few data can help us here but Figure~4 in \citet{cole1998-loadingIJSS} shows model predictions for the effective modulus at four loading frequencies that include those associated with surface gravity wave periods, i.e.\ $10^{-2}$--$10^0$\,Hz (or 0.01 -- 1 Hz), and, incidentally, the reduction in $Y$ due to total porosity, i.e.\ air plus brine. The latter effects are comparable in magnitude to the reductions in $Y$ given above; the effect of rate is about 0.5\,GPa as wave period is changed from 1\,s to 10\,s, and about 1\,GPa from 10\,s to 100\,s. We therefore consider a reduction of 1\,GPa is to be reasonable in our model, and in summary
\begin{equation}\label{Y-eff-formula}
Y^{*}=Y_0(1-3.51\ups_b)-1\textrm{GPa}.
\end{equation}
This is plotted in Figure~\ref{fig-iceprops}(b). %{\bf**comment on values?**}
An appropriate choice of a value for the effective Young's modulus is important from the wave modeling perspective too, as the higher $Y^*$ becomes the more energy is reflected at each floe present and the greater the attenuation experienced by the wave train. However, because the same value of $Y^*$ is used to convert from flexural stress to failure strain, the analysis is self-consistent.

The final property we will need to consider is Poisson's ratio. \citet{langleben-pounder1963} determined it to be $\nu = 0.295\pm 0.009$ from seismic measurements, so in most wave calculations involving ice \cite[e.g.][]{fox_squire91} it is simply taken to be 0.3. In Figure~\ref{fig-iceprops}(c), we plot the breaking strain using this value of Poisson's ratio, combined with (\ref{sig-c-formula}) and (\ref{Y-eff-formula}) in
\begin{equation}\label{eps-c-formula}
\eps_{c}=\frac{\sig_{c}}{Y^{*}(1-\nu^{2})}.
\end{equation}
The breaking strain has a minimum value of approximately $5.3\times10^{-5}$ when $\ups_{b}=0.15$ ($Y^{*}=3.8\,$GPa). The value is approximately constant for $\ups_{b}\in[0.1,0.2]$. It shows an increase for both higher and lower brine volumes --- the less porous ice is predictably stronger, while the more porous ice is more compliant so will be able to sustain more bending before breaking. If $\ups_{b}=0.05$, $\eps_{c}\approx7.1\times10^{-5}$ ($Y^{*}=7.2\,$GPa), while if $\ups_{b}=0.1$, the breaking strain drops to $\eps_{c}\approx5.5\times10^{-5}$ ($Y^{*}=5.5\,$GPa). If Poisson's ratio $\nu=0$, these values fall by about 9\%, so that the minimum is $4.8\times10^{-5}$.

\subsection{Fatigue}\label{sec-fat}
The above discussion of the anelastic response of sea ice does not preclude the possibility that floes can gradually fatigue due to repeated bending imposed by passing waves. Fatigue, whether of the high-cycle type associated with elastic behavior and growth of microscopic cracks that eventually reach a critical size for fracture, or low cycle fatigue where the stress is sufficient for plastic deformation, is characterized by cumulative damage such that materials do not recover when rested, i.e.\ they behave inelastically as opposed to anelastically. Accordingly, the effective modulus approach described above, which includes only fully recoverable elastic deformation, cannot accommodate fatigue. There is, however, a suggestion \citep{langhorne-etal1998} that an endurance limit, i.e.\ a value of stress for which a material will retain its integrity even when subjected to an infinite number of load cycles, exists for sea ice.  This value, 0.6, was determined on stationary shore fast sea ice in McMurdo Sound, Antarctica. \citet{dumont_etal2011} therefore reduced their flexural strength by a factor of 0.6. We, on the other hand, have chosen not to do this because (i)~the ice and wave conditions change rapidly in the MIZ so, while a stress greater than $0.6\sig_{c}$ can cause failure in principle, it may still occur at a timescale that is well beyond that associated with the local dynamics (recall that the endurance limit is for infinite time), (ii)~fatigue strictly negates the use of an effective modulus, as permanent irrecoverable damage is gradually done to the sea ice either by the nucleation and propagation of cracks or by secondary and tertiary creep, and (iii)~the fast ice data of \citet{langhorne-etal1998} show considerable scatter, which is a common feature of fatigue experiments even for simple materials. We rest content, therefore, with the expression for $Y^{*}$ defined in equation (\ref{Y-eff-formula}),
noting that fatigue can easily be added at a later point if results indicate that it plays a role.
%noting that fatigue can be added subsequently if required with little work.

\subsection{Attenuation models}\label{sec-atten}
The operational model selects the rate of exponential attenuation from a pre-calculated look-up table. This action is performed for each grid cell and at each time step and depends on the prevailing wave and ice conditions. The look-up table describes the functional dependence of the attenuation rate on the properties of the ice cover and the wave field. The attenuation model runs independently of the operational model.

In this section three attenuation models will be discussed. They are all based on linear wave scattering theory. The ice is represented by a thin-elastic plate and only responds to fluid motion in flexure, with its position on the surface of the ocean fixed. Wave energy may propagate freely through homogeneous regions of the ice-covered ocean. Attenuation is not caused by the loss of energy from the fluid-ice system but is due to energy being scattered by imperfections in the ice cover. Motion is confined to two-dimensional transects, i.e.\ one horizontal dimension and one depth dimension. Scattering, and hence attenuation, is produced solely by floe edges. Extensions to scattering by other features in the ice cover, such as cracks and pressure ridges, are possible \cite[see][]{bennetts-squire2011-atten} but the restriction to floe edges is sufficient for the present investigation. The fluid surface is perforated with a large number of floes and an incident wave originates at one spatial extreme. Assuming time-harmonic conditions of a given period, the profile of the wave in the transect is calculated by solving the full multiple scattering problem, i.e.\ the infinite sum of reflections and transmissions of the wave between floe edges. Exponential decay is then a product of localization theory, which relies on positional disorder and requires proper consideration of wave phases. The reliance on disorder implies the use of an averaging approach. The attenuation coefficient, $\alpha$, is hence calculated as the ensemble average of the attenuation rates produced in simulations that are randomly selected from prescribed distributions. It is natural to calculate a non-dimensional attenuation coefficient, i.e.\ per floe, for these types of problem, but this is easily mapped onto a dimensional attenuation coefficient, i.e.\ per meter, for use in the operational model.

The attenuation models now differ only in the distribution of floes that they use, although they may also incorporate certain approximations to facilitate the solution process. The first model considered, denoted A, uses a seemingly plausible choice for the distributions. The floe length distribution is based on the split-power law discussed in \S\ref{sec-fsd} and given in (\ref{pdf-split-power-law}) with $\bfD=(20$\,m,\,$D_{c})$, $\bs\g=(1.15,2.5)$ and $\prob_{0}$ chosen to make the distribution continuous. It was derived from observations in Antarctic locations. Floe separations are generated from an exponential distribution $\prob(G > g) = \exp{(-g/\langle G\rangle)}$, with $\langle G\rangle = \langle D\rangle(c^{-1} - 1)$ and ice concentration $c = 0.9$. \citet{toyota-etal2011-floedist} did not measure a floe separation distribution so we picked this distribution arbitrarily. The attenuation coefficient is calculated as the average of 100 randomly generated simulations.

The second and third models are based on the recent work of \citet{bennetts-squire2011-atten}. Rather than considering spatial distributions, \citet{bennetts-squire2011-atten} considered the wave phases as random variables and averaged over all possibilities, assuming they were uniformly distributed. 
%This approach is consistent with the basis of random linear wave theory \cite[see, e.g.][]{krogstad2005-dirspec},
%and was adopted because of the limitations of the two-dimensional model. 
They argued that the model is not intended as a true replica of the MIZ but rather a series of impedance changes that resemble the transition a wave experiences when passing between open water and an ice-covered region of the ocean.
Consequently, detailed predictions about the exact distribution of wave phases cannot be relied upon, and an assumption of uniformity is the simplest possible in the absence of a more realistic model.
In this setting the attenuation coefficient may be calculated analytically rather than relying on a numerical approximation. The expression for the attenuation coefficient can be simplified further if the floes are assumed to be long, so that only the reflection produced by a single floe edge is required, and the attenuation coefficient is then given by
$\a=-2\log(1-\vert R\vert^{2})$, where $R$ is the scattering coefficient by a floe edge of the specified thickness. Model B uses attenuation calculated in this manner.

In several previous investigations it has been identified that scattering models under-predict attenuation for large periods. This may be a result of the increased influence of unmodeled dissipative mechanisms, although it is unclear which, if any, is dominant in this regime. One plausible candidate is that some secondary creep is occurring at longer periods where the flexural straining rates are slower. While this issue remains unresolved it is useful to add an artificial viscosity to the scattering model to dampen the large period waves. \citet{bennetts-squire2011-atten} showed that their model could be modified in a simple manner to accommodate viscosity in the form of an imaginary term in the fluid-ice coupled dispersion relation. Modifying the attenuation model B in this way gives a new model, C, and the value of the damping parameter was set by matching results to the most reliable MIZ experimental data available at present \citep{squire_moore80}.

Another failing of all the models is that they don't predict the maxima in the attenuation coefficients (rollover) observed by, e.g.\ \citet{wadhams-etal1988-atten}, below wave periods of about 5--6\,s. It has been suggested that this could come from additional local wave generation, possibly caused by nonlinear wave generation \citep{wadhams_etal86-directional-spectrum}. However, we will not address this issue further in this paper, which unfortunately could mean that we are underestimating the extra strains produced by additional wave energy at low periods. The three models are summarized in Table~\ref{tab-atten}. Note that all three models include the draft of the floes, which alters floe scattering in some circumstances \citep{williams-squire2010-thickness} potentially affecting results significantly. Draft is actually computationally important as well, because it removes artificial resonances and thus makes averaging over thicknesses redundant. Consequently, the neglect of draft is considered to be ill-advised.
\begin{table}[htdp]
\begin{center}
\begin{tabular}{l c c c}
\hline
Attenuation Model & A & B & C\\
\hline
Draft & Yes & Yes & Yes\\
Uniform Phases & No & Yes & Yes\\
%Averaging Method & Geometric & Geometric & Geometric\\
Calculation Method & Numerical & Analytical & Analytical\\
Damping & No & No & Yes\\
\hline
\end{tabular}
\end{center}
\caption{Attributes of the different attenuation models used in the paper. \textit{Draft} refers to whether or not draft is included in the scattering model; \textit{Uniform Phase} refers to whether or not the phases of the waves incident at each edge are uniformly distributed or whether they are associated with the floe length distribution;
\textit{Calculation Method} refers to whether the ensemble averaging is done numerically, or using an analytic formula; and \textit{Damping} refers to whether additional attenuation is added to the attenuation predicted by the multiple scattering model.}
\label{tab-atten}
\end{table}%

\section{Waves-in-ice models (WIMs)}\label{sec-WIMs}
In our results section we will compare the effects of three WIMs. These involve different methods of including the effects of waves in the MIZ and different combinations of breaking criteria. {WIMs\,1 and 2} are based on the approach of \citet{dumont_etal2011}, and treat different frequencies as independent wave groups that separate out as they travel due to dispersion. For direct comparison with \citet{dumont_etal2011}, {WIM\,1} uses both stress and strain breaking criteria, although, as foreshadowed in \S\,1, we do not advocate using a stress criterion because of the unrealistic breaking it predicts when the waves are long. Although {WIM\,1 is} similar to the model of \citet{dumont_etal2011}, certain
changes have been made to facilitate comparison with {WIMs\,2 and 3}.
{WIM\,2} is identical to {WIM\,1}, except that it only uses a strain breaking criterion.

The final model, {WIM\,3}, has the fundamental difference of allowing  interactions between different frequencies. This requires the consideration of integrals (approximated numerically) over the entire wave spectrum. This approach is similar to those used by \citet{rottier1992-jgr-floepairs} to deal with floe collisions and accoustic noise production, and \citet{langhorne-etal1998}, \citet{kohout_meylan08} and \citet{vaughan-squire2011-fracprob} who also deal with ice break-up. {WIM\,3}  only uses a strain breaking criterion.

Before we can describe the WIMs, however, we first introduce the approximations we use for the stresses and strains in the ice floes.

\subsection{Strain approximation}\label{sec-strain}
Let us assume that a sinusoidal wave profile
\begin{equation}\label{w-cos}
w_\textrm{ice}(x,t)=\tilde A\cos(k_{\mathrm{ice}}x-\om t)
\end{equation}
with period $T=2\pi/\om$ is traveling in the ice, where
$\tilde A=AW(\om)$ and the factor $W(\om)=k_{\mathrm{ice}}/k_{\mathrm{water}}$ converts the wave amplitude for open water $A$ into a wave amplitude for ice-covered water. The quantity $k_{\mathrm{water}}=\om^{2}/g$ is the infinite depth wave number for open water, and $k_{\mathrm{ice}}$ its analogue for ice-covered {water found} from (\ref{disprel}) in Appendix~\ref{app-disprel}. We note that $k_{\mathrm{ice}}$ can be approximated by the open water wave number $k_{\mathrm{water}}=\om^{2}/g$ in some circumstances, as was done by \citet{dumont_etal2011}, making $W\approx1$.
However, the approximation is not used in the current paper, as it affects the FSDs predicted by the model. Note that it mainly affects the lengths of floes rather than the MIZ width itself.

Now, using the formula (\ref{eps-max-app}) from {Appendix~\ref{app-EBplate},} the maximum strain in the ice can be approximated by
\begin{align}
\eps_{\max}
&=\frac{k^{2}_{\mathrm{ice}}h}{2}W(\om)A=E(\om)A,
\end{align}
where $E(\om)$ is the strain per metre of wave amplitude.
The spectral density function for the strain in the ice will then be $S(\om)E^{2}(\om)$, where $S$ is the wave spectrum that produces the open water displacement.

\subsection{Stress/pressure approximation}\label{sec-stress}
\citet{dumont_etal2011} hypothesize that cavitation between the ocean surface and the underside of an ice floe would result in a pressure loading that could cause the floe to break. The model they envision replaces the ice floe with an elastic beam of length $L$ that rests horizontally on a sinusoidal wave profile centered in the trough between two adjacent crests, so that $L=\lambda_\mathrm{ice}$. Thus breaking only occurs at wave troughs. We, however, consider a beam of length $L=\lambda_\mathrm{ice}/2$, supported at
the mean ice-water interface (i.e.\ between troughs and crests), so that forces on the beam are maximal (and approximately equal) at both crests and troughs. These forces are approximated here in the same way as by \cite{dumont_etal2011}, and are taken to be the average of the pressures (integrated over the whole beam) due to buoyancy and gravity [see {Figure~}4 of \citeauthor{dumont_etal2011}, \citeyear{dumont_etal2011}]. This implies that
\begin{equation}\label{Q-sig}
Q(\om)A=\frac1\pi\langle\rho\rangle g\lambda_\mathrm{ice}W(\om) A,
\end{equation}
where $Q(\om)$ is the force per unit beam width per metre of wave amplitude, and $\langle\rho\rangle=(\rho_{\mathrm{ice}}+\rho_{\mathrm{water}})/2$.

The maximum possible value of $QA$ that a beam can endure, which we shall denote by $Q_{c}$, can be approximated by comparing the wave profile to a three point beam-loading setup, where the load acts halfway between the  supports with a force $F$. The maximum force that the beam can withstand without breaking is denoted by $F_{c}$. From Euler-Bernoulli beam theory \citep[see][and Appendix~\ref{app-3pt}]{schwarz-etal1981-icestrength},
$F_{c}$ is related to the flexural strength $\sig_{c}$ by
\begin{equation}\label{sigc-3pt-txt}
\sig_{c}=\frac{3F_{c}L}{2bh^{2}}=\frac{3Q_{c}\lambda_\mathrm{ice}}{4h^{2}},
%=\frac{12F_{c}Lh}{8bh^{3}}
%=\frac{F_{c}Lh}{8I}
\end{equation}
where $b$ is the beam width, $h$ is the beam thickness and $Q_{c}=F_{c}/b$. At this
point we make one further adaptation to the model of \cite{dumont_etal2011}, as ice is better represented by a plate model than a beam model. In Appendix~\ref{app-EB} we discuss why it is more difficult to bend a plate than a beam, and thus a larger stress is needed to make the strain exceed a certain breaking strain in a plate than in a beam. Therefore (\ref{sigc-3pt-txt}) should be modified by increasing $\sig_{c}$ to $\sig_{c}^{*}=\sig_{c}/(1-\nu)^{2}$ and the maximum force per unit width that a plate can withstand is
\begin{equation}\label{Qc-3pt}
Q_{c}=\frac{4h^{2}\sig^{*}_{c}}{3\lam_{\mathrm{ice}}}.
\end{equation}

\subsection{Incident waves}\label{sec-inc}
Most wave data are given parametrically in terms of the significant wave height $H_{s}$, the peak period $T_{\rm M}$ and the mean wave direction \citep{ochi1998-book-waveprobs,wmo1998-wave-forecast}. However, in our one-dimensional simulations we will only use $H_{s}$ and $T_{\rm M}$, and take the waves to be traveling directly into the ice. The significant wave height here is related to the variance of the sea surface displacement (see below).
For the power spectrum of the incident wave we invoke the Bretschneider spectral density function (SDF) $S_{\rm B}$ given by
\begin{equation}\label{sdf-bret}
S_{\rm B}(\om;T_{\rm M},H_{\rm s})=\frac{1.25H_{\rm s}^{2}T^{5}}{8\pi T_{\rm M}^{4}}
\exp\left(-\frac{5T^{4}}{4T_{\rm M}^{4}}  \right).
%\rme^{-1.25(T/T_{\rm M})^{4}},
\end{equation}
The incident wave spectrum will be moved into the ice using the Advection-Attenuation Schemes (AASs) described in \S\ref{sec-adv}, so that we will always be able to determine the spectral density function (SDF) $S(\om)$ at any point in the ice field.

\subsection{Waves-in-ice models 1 and 2}\label{sec-WIM12}

\citet{dumont_etal2011} argue that, as a result of both attenuation and dispersion, wave groups of different frequency will separate out and the effects of each group would be able to be considered in turn (beginning with the faster, lower frequency groups). Integrating the SDF $S(\om)$ with respect to $\omega$ provides the mean square value of the sea surface elevation $\langle w^{2}\rangle$, and also the mean square wave amplitude $\langle A^{2}\rangle$, since it is commonly assumed that $\langle w^{2}\rangle=\frac12\langle A^{2}\rangle$. The SDF is a function of the frequency that is built from a random phase-amplitude assumption \citep{holthuijsen2007-wavesbook}. However, using the SDF to produce an amplitude spectrum $\mathscr{A}(\om)=\langle A^{2}(\omega)\rangle^{1/2}$ as a function of frequency is not straightforward. One method would be to {integrate} over a frequency interval $\left[ \omega - \frac12\Delta\omega, \omega + \frac12\Delta\omega \right]$. Unfortunately, $\mathscr{A}(\omega)$ then depends on the arbitrary choice of $\Delta\omega$, which is not ideal for the purpose of comparing different waves-in-ice models. In order to rule out this {arbitrariness}, {\citet{dumont_etal2011}} used a simple dimensional analysis where they considered $[S]=[\frac{1}{2} \mathscr{A}^2/\om]$ to obtain
\begin{equation}\label{Aexp-wgm}
\mathscr{A}(\omega) \simeq \frac{k_{\mathrm{ice}}}{k_{\mathrm{water}}}\sqrt{2\om S(\om)},
\end{equation}
where we have {included} the correction for ice-coupled waves. Despite its unconventional approach, this formula does appear to provide a reasonable and conservative estimate of the highest waves of a group although it tends to overestimate the amplitude of short waves; the ones that are the most rapidly attenuated. An alternative formulation based on spectral integrals is presented later to resolve this issue and quantify the uncertainty. Comparison between both formulations show similarities well within our remaining uncertainties.

Having obtained an expression for the expected wave amplitude, the stress and strain amplitudes are calculated {and, if} they exceed $\sig_{c}$ or $\eps_c$ respectively, the ice is {broken.}
(As discussed in \S2.3, \citet{dumont_etal2011} also multiplied $\sig_{c}$ by a fatigue coefficient $\mu=0.6$, but this approach is not adopted here.)
Thus, the ice is broken if either of the following two conditions are satisfied:
\begin{subequations}
\begin{align}
\mathscr{ A}(\om)&>A_{c}^{\sig}=\frac{ Q_{c}}{ Q(\om)}
=\frac{{4}\pi h^{2}\sig^{*}_{c}}{3\langle\rho\rangle g\lam^{2}_{\mathrm{ice}}(\om)W(\om)},\label{stress-con}\\
\mathscr{ A}(\om)&>A_{c}^{\eps}=\frac{\eps_{c}}{E(\om)}
=\frac{2\eps_{c}}{k_{\mathrm{ice}}^{2}hW(\om)}=\frac{\lam_{\mathrm{ice}}^{2}\eps_{c}}{2\pi^{2}hW(\om)}.\label{strain-con-wgm}
\end{align}
\end{subequations}
If breaking occurs for a particular frequency, the maximum floe size allowed in a particular grid cell is changed to $\lam_\mathrm{ice}/2$.
As noted above, it is questionable whether (\ref{stress-con}) should be included at all for long waves, as it implies that very long waves need only a very small amplitude to break ice. Such waves have very low curvature so ice floes are easily able to bend and follow their profile, which challenges the fundamental assumptions made in deriving (\ref{stress-con}). Nonetheless, an alternative method of including the effects of cavitation and wetting, or other breaking mechanisms, {would be} worth considering in the future, but such methods should reflect the fact that breaking events are more likely to be caused by shorter waves than longer ones. Despite the misgivings about the stress failure condition (\ref{stress-con}), for comparision with \citet{dumont_etal2011} it is useful to include it in {WIM\,1}.

Summarizing, we now reiterate the three WIMs we will be using.
{WIMs\,1 and 2}, like \citet{dumont_etal2011}, both {consider wave groups individually}, using (\ref{Aexp-wgm}). In {WIM\,1}, the ice is broken if either (\ref{strain-con-wgm}) or (\ref{stress-con}) are satisfied, while in {WIM\,2}, the ice is broken only if (\ref{strain-con-wgm}) is satisfied. In the next section {WIM\,3} is introduced{, which}, like {WIM\,2}, only applies a strain criterion for ice breaking, but it allows different frequencies to interact. It turns out that, while less orthodox in its use of wave statistics derived from the spectral density function than {WIM\,3}, {WIM\,2} produces similar results.

\subsection{Waves-in-ice model 3}\label{sec-WIM3}
In this model we allow for the interaction between waves of different frequencies by integrating over the entire wave spectrum. The mean square sea surface elevation, or the variance in the position of a water particle at the sea surface, $\langle w^{2}\rangle=m_{0}[w]$ can be obtained from the formula
\begin{equation}
m_{n}[w]=\int_{0}^{\infty}\om^{n}S(\om)\rmd\om,
\end{equation}
where $S(\om)$ is the SDF at a general point in space; either in the open ocean or within the ice after having undergone some attenuation.
(We will also use the second moment of the SDF, $m_{2}$, later on.) The significant wave height is related to $m_{0}$ by $H_{s}=4\sqrt{m_{0}}$.

Wave heights generally follow a Rayleigh {distribution, for which the probability} of a wave amplitude exceeding a certain value $A_{c}$ is approximately
\begin{align}\label{rayleigh-wave-prob-orig}
\prob(A>A_{c})&=\rme^{-A^{2}_{c}/\langle A^{2}\rangle},
\end{align}
\citep{longuet-higgins1952-rayl,longuet-higgins1980-rayleigh}, where $\langle A^{2}\rangle$ denotes the mean square amplitude. If the wave spectrum has a narrow bandwidth and non-linear effects are negligible (low wave steepness), then
$\langle A^{2}\rangle=2m_{0}[w]$, so
\begin{align}\label{rayleigh-wave-prob-lin}
\prob(A>A_{c})&=\rme^{-A^{2}_{c}/2m_{0}[w]}.
\end{align}
\citet{forristall1978-rayleigh} showed that some data were not fitted well by (\ref{rayleigh-wave-prob-lin}),
while in response \citet{longuet-higgins1980-rayleigh} showed that these data were fitted well by (\ref{rayleigh-wave-prob-orig}), putting the difference down to non-linear effects.

The mean square ice displacement is approximately
\begin{equation}
\langle w_\textrm{ice}^{2}\rangle=m_{0}[w_\textrm{ice}],\ \
m_{n}[w_\textrm{ice}]=\int_{0}^{\infty}\om^{n} S(\om)W^{2}(\om)\rmd\om.
\end{equation}
In a similar way to the water wave spectrum, the probability of the amplitude $\tilde A$ exceeding a certain value $A_{c}$ is $\prob(\tilde A>A_{c})=\exp(-A_c^{2}/2m_{0}[w_\textrm{ice}])$.
In addition, we can also estimate the number of waves we expect in a time interval
$\D t$, $N_{\rm W}$, as
\begin{align}\label{Ndownzeros-WMO}
N_{\rm W}=\frac{\D t}{2\pi}\sqrt{\frac{m_{2}[w_\textrm{ice}]}{m_{0}[w_\textrm{ice}]}}
\end{align}
\citep{wmo1998-wave-forecast}. More precisely, this is the number of times we can expect a particle to cross its point of mean displacement in a downward direction. Note that equations (\ref{rayleigh-wave-prob-lin}) and (\ref{Ndownzeros-WMO}) disagree with the results presented by \citet{cartwright-longuet1956-rayleigh}. Presumably they are typographical errors, as \citet{longuet-higgins1980-rayleigh} uses equation (\ref{rayleigh-wave-prob-orig}).
$N_{\rm W}$ also defines a representative wave period for the spectrum $S$ at a given point of $T_{\rm W}=\D t/N_{\rm W}$, and a representative wavelength (in ice) of $\lam_{\rm W}$ determined from (\ref{disprel}). Inside the ice cover, $T_{\rm W}$ plays a similar role to $T_{\rm M}$, the peak period parameter in the Bretschneider spectrum.

We can also define an analogous quantity for the strain. Its mean square value is
\begin{align}
\langle\eps^{2}\rangle=m_0[\eps],\ \
m_n[\eps]
=\int_{0}^{\infty}\om^{n} S(\om)E^{2}({\om})\rmd\om,
\end{align}
so $\prob_{\eps}=\prob(E_{\rm W}>\eps_{c})=\exp(-\eps_c^{2}/2m_{0}[\eps])$, where $E_{\rm W}$ is the maximum strain induced in a floe by a passing wave. The probability that none of the $N_{\rm W}$ waves will cause $E_{\rm W}$ to exceed $\eps_{c}$ is then $(1-\prob_{\eps})^{N_{\rm W}}$. Thus to determine if the ice will break we will require that
\begin{equation}\label{strain-con0}
1-(1-\prob_{\eps})^{N_{\rm W}}>\prob_{c},
\end{equation}
or, equivalently, that $\prob_{\eps}>\tilde\prob_{c}=1-(1-\prob_{c})^{1/N_{\rm W}}$. If it is, {we then decrease} $D_{\rm max}$ to $\lam_{\rm W}/2$ (unless $\lam_{\rm W}/2<D_{\rm min}$). The probability threshhold $\prob_{c}$ is flexible, but $\prob_{c}=0.5$ is a reasonable choice since we only have two possible outcomes (breaking or not breaking). If we now define a significant strain amplitude $E_{s}=2\sqrt{m_{0}[\eps]}$, (\ref{strain-con0}) can be written in terms of $E_{s}$ as
\begin{equation}\label{strain-con}
E_{s}>E_{c}=\eps_{c}\sqrt{{-2}\big/{\log\big(\tilde\prob_{c}\big)}},
%\big(-0.5\log\big[1-(1-\prob_{c})^{1/N_{\rm W}}\big]\big)^{-1/2},
\end{equation}
which shows more explicitly how the strain criterion depends on $\eps_{c}$, $\prob_{c}$ and $N_{\rm W}$. We will return to the sensitivity of our results to these parameters in \S\ref{sec-testsens}.

\subsection{Advection and attenuation}\label{sec-adv}

We have two AASs that we will use to move the waves from the open ocean into the ice. As we shall see, which one we use turns out to have an extremely large effect on our results. The first one, which we shall label AAS\,1, is similar to the one introduced by \citet{dumont_etal2011}. The difference is that it advects the spectral density function $S$ itself, as opposed to the mean square amplitude $\mathscr{A}$. The scheme proceeds as follows. Let us discretize our space, time and frequency variables using $x_{j}=j\D x$ ($j=0,1\ldots$), $t_{n}=n\D t$ ($n=0,1\ldots$), and $\om_{k}=\om_{0}+k\D\om$ $k=0,1,\ldots,N_{\om}$. We choose $\om_{0}$ and $\D\om$ so that periods between 2.5\,s and 25\,s are included. Then let $s_{k}=a_{k}\D t$ be the distance a wave moves each time step. The frequency-dependent group velocity $a_{k}$ should depend on space also but this proved very difficult to implement successfully due to the very large contrast between the wave speeds in ice and water for high frequencies. Then the Courant-Friedrichs-Lewy (CFL) number, i.e.\ the fraction of a grid cell this distance corresponds to, is $\Cop_{k}=s_{k}/\D x$. Like the group velocity, the CFL number is also frequency-dependent. It should be less than or equal to 1 to avoid numerical diffusion due to waves crossing more than one grid cell each time step, and also aliasing between waves traveling in opposite directions. As a final notational shorthand let $S_{k,i}^{n}=S(\om_{k},x_{i},t_{n})$.
Then the AAS proceeds as follows:
\begin{subequations}
\begin{align}
\mathscr{S}_{k,j}^{n}&=S_{k,j}^{n-1}
%\notag\\&\quad
+\Cop_{k}\left(  S_{k,j-1}^{n-1}-S_{k,j}^{n-1}\right),\\
\hat \a_{k,j}^{n-1}&=\frac{c_{j} \a_{k,j}^{n-1}}{\langle D_{j}^{n-1}\rangle},\label{alp_dim}\\
S_{k,j}^{n}&=\mathscr{S}_{k,j}^{n}\exp\left(-\hat\a_{k,j}^{n-1}s_{k}\right).
\end{align}
\end{subequations}
Here, $\hat\a_{k,j}^{n}$ is the dimensional attenuation coefficient that depends on the average floe length $\langle D_{j}^{n}\rangle$ and ice concentration $c$, and is updated every time step after breaking. It is a well-known phenomenon that using a CFL number that is too much less than 1 can lead to significant numerical damping of waves.
However, an additional complication in a situation where physical attenuation also occurs is that
when $\Cop_k<1$ this attenuation is numerically magnified.
%For example, i
In the case where $\Cop_{k}=1$, the wave arrives in a grid cell, {possibly breaks {the ice in that cell},} and then leaves the grid cell. However if $\Cop_k<1$, if some breaking has been done and the attenuation (in the entire cell) increased, the wave still has to cross {a proportion of the} {cell} {before it is advected to the adjacent cell}.
Thus{,} due to its slightly slower speed, it has significantly less energy than it would have had if it had crossed a whole grid cell each time step.

To correct for this, in our second advection-attenuation scheme, {AAS\,2}, we introduce a frequency-dependent time step $\tau_{k}=\D x/a_{k}\geq\D t$, so each wave moves a distance $\D x$ each time step (the CFL number $\Cop_k=1)$. Now, define time indices ${l_{k,m}}$ and $u_{k,m}$ so that if $l_{k,m}\leq n\leq u_{k,m}$, $t_{n}\in\mathscr{I}_{k,m}=\big[(m-1)\tau_k,m\tau_k]$. In other words, $n={l_{k,m}}$ represents the first time $t_{n}$ moves into the time interval $\mathscr{I}_{k,m}$, and $n={u_{k,m}}$ represents the last time it does so. Moreover,  because we have defined $\Cop_{k}=1$ for all $k$, moving from one interval $\mathscr{I}_{k,m}$ to another is equivalent to moving from one grid cell to another. Although the spatial intervals, or grid cells, are more easily {envisioned} than the temporal intervals $\mathscr{I}_{k,m}$, they are unfortunately more difficult to program as this involves keeping track of individual wave packets. The crux of AAS\,2 is that just {before waves of} a particular frequency change grid cell, the attenuation coefficient for those frequencies in the next grid cell is fixed until {the waves leave the grid cell beyond,} and so only the waves traveling behind them experience the increased attenuation due to any breaking that they do.

\begin{table}[h]
\caption{Default values for independent model parameters.  FSD stands for floe size distribution, and $\prob_{\max}=\prob(D<D_{\max})$ is the proportion of floes taken to have diameters below $D_{\max}$.
If $D_{\rm max}>D_{\rm unif}$, then the floe lengths in a cell are all $D_{\max}$.
}
\begin{center}
\begin{tabular}{l c c}
\hline
Quantity & Symbol & Value\\
\hline
Ice thickness & $h$ & 2\,m\\
Ice concentration & $c$ & 0.95\\
Water density & $\rho_{w}$ & 1025\,kg\,m$^{-3}$\\
Ice density & $\rho_{i}$ & 922.5\,kg\,m$^{-3}$\\
Gravitational acceleration & $g$ & 9.81\,m\,s$^{-2}$\\
Poisson's ratio & $\nu$ & 0.3\\
Brine volume fraction & $\ups_{b}$ & 0.1\\
Exponent for small floe FSD & $\g_{1}$ & 1.15\\
Exponent for large floe FSD & $\g_{2}$ & 2.5\\
Minimum floe size in FSD & $D_{\min}$ & 20\,m\\
FSD cut-off length  & $D_{\rm unif}$ & 200\,m\\
Spatial resolution & $\D x$& 5\,km\\
Time step & $\D t$& 400\,s\\
Minimum wave period & $2\pi/\om_{30}$ & 2.5\,s\\
Maximum wave period & $2\pi/\om_{0}$ & 23.8\,s\\
Spectral resolution & $\D \om$ & $7.5\times10^{-2}$\,s$^{-1}$ \\
Breaking probability threshold & $\prob_{c}$&0.5\\
$\prob(D<D_{\max})$ & $\prob_{\max}$ & 0.95 \\
\hline
\end{tabular}
\end{center}
\label{tab-indepvals}
\end{table}%

Bearing the above in mind as we return to our algorithm, we first calculate the spectral density at $t=(m-1)\tau_{k}$ from
\begin{subequations}
\begin{align}
\hat\a_{j,k}^{m-1}&=\frac{c_{j}\a_{j,k}}{\big\langle{D}^{u_{k,m-2}}_{j}\big\rangle},\\
%\sum_{n=n_{0}'}^{n'_{1}}\Big(\overline{D}_{i}(t_{n})  \Big)^{-1},\\
%F_{j,k}^{m-1}&=S(\om_{k},x_{j},(m-1)\tau_{k})
F_{j,k}^{m-1}&=S\big(\om_{k},x_{j},(m-1)\tau_{k}\big)
\notag\\&
=\mathscr{F}_{j,k}^{m-1}\exp\left(-\hat\a_{j,k}^{m-1}\D x\right),
\end{align}
\end{subequations}
where $\hat\a_{j,k}^{m-1}$ now represents the attenuation when $t=(m-1)\tau_{k}$, and $\mathscr{F}_{j,k}^{m-1}$ has been calculated already. As mentioned above, to stop the waves getting trapped in the cell, the next time the attenuation gets updated is at $t=m\tau_{k}$. We now calculate the unattenuated spectral density at $t=m\tau_{k}$ by advecting the waves using the $\tau_{k}$ time step, which is designed to be more (perfectly) efficient. This gives us:
\begin{align}
\mathscr{F}_{j,k}^{m}&=F_{j-1,k}^{m-1}.
\end{align}
We can now interpolate between $t=(m-1)\tau_{k}$ and $t=m\tau_{k}$ to find the spectral density at $t=t_{n}$ using:
\begin{subequations}
\begin{align}
s_{k}&=a_{k}\big(t_{n}-(m-1)\tau_{k}\big),\\
\mathscr{S}_{j,k}^{n}&=F_{j,k}^{m-1}
+\frac{s_{k}}{\D x}\left(\mathscr{F}_{j,k}^{m}- F_{j,k}^{m-1}\right),\\
S_{j,k}^{n}&=\mathscr{S}_{j,k}^{n}\exp\left(-\hat\a_{j,k}^{m-1}s_{k}\right).
\end{align}
\end{subequations}
If we are using {WIM\,1 or 3}, having $S_{j,k}^{n}$ now lets us calculate $\mathscr{A}$ from (\ref{Aexp-wgm}) and thus determine the new floe size distribution. Similarly, with {WIM\,3} we can approximate the integrals $m_{0}[w_\textrm{ice}]$, $m_{2}[w_\textrm{ice}]$ and $m_{0}[\eps]$, which are required to determine if breaking occurs in that context.

\section{Results}
We begin our results section by comparing the different attenuation models. As model A is found to give unphysical results, we concentrate on the best two attenuation models and run some idealized simulations with constant ice concentration and idealized thickness profiles to test the sensitivity of our results to the various model parameters. Finally we run some realistic simulations using 2007 wave and ice model data from the 79$^{\circ}$N transect of the Fram Strait between 3$^{\circ}$E (where the waves are specified) and where the transect hits the island Norske \O{er} (off the east coast of Greenland) at longitude 17.7$^{\circ}$W.

\begin{table}[h]
\caption{Default values for dependent model parameters. FSD stands for floe size distribution.}
\begin{center}
\begin{tabular}{l c c c}
\hline
Quantity & Symbol &  Dependencies & Value\\
\hline
Flexural strength & $\sigma_{c}$ & $\ups_{b}$  & 0.27\,GPa\\
Effective Young's modulus & $Y^{*}$ & $\ups_{b}$ & 5.5\,GPa\\
Breaking strain & $\eps_{c}$ & $\sigma_{c}, Y^{*}, \nu$ &$4.6\times10^{-5}$\\
Fragility & $\Pi$ & $\g_{1}$ & 0.55\\
Small floe cutoff in FSD & $D_{c}$ & $Y^{*}, h, \nu, \rho_{w}, g$ & 54.6\,m \\
\hline
\end{tabular}
\end{center}
\label{tab-depvals}
\end{table}

Tables~\ref{tab-indepvals}, \ref{tab-depvals}, and \ref{tab-timevals} list the values used for the model parameters. In all simulations, unless {otherwise} specified, we take the wave speeds to be spatially invariant, and for all frequencies to have CFL value $\Cop_{k}=1$. In this case {AAS\,1} and {AAS\,2} are equivalent.

\begin{table}[h]
\caption{Model variables that are updated each time step in the model. $\prob_{0}=\prob(D<D_{c})$ is the proportion of floes that are small.}
\begin{center}
\begin{tabular}{l c c}
\hline
Quantity & Symbol & Initial Value\\
\hline
Maximum floe length in grid cell& $D_{\max}$ & 500\,m\\
Mean floe length in grid cell& $\langle D\rangle$ & 500\,m\\
Maximum floe length in MIZ & $D_{\miz}$ &0\,m\\
Width of MIZ & $L_{\miz}$ &  0\,km\\
$\prob(D<D_{c})$ & $\prob_{0}$ & 0\\
Significant strain & $E_{s}$& 0\\
Expected number of waves & $N_{\rm W}$ & 0 \\
\hline
\end{tabular}
\end{center}
\label{tab-timevals}
\end{table}

\subsection{Attenuation results}

\begin{figure}[ht]
\begin{center}
\hspace{-25mm}\includegraphics[width=108mm]{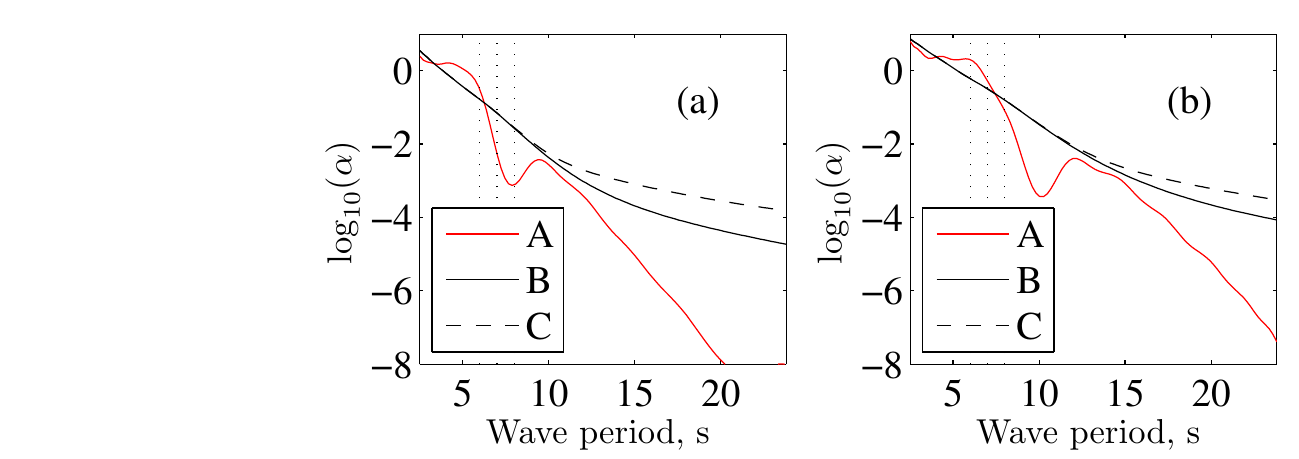}
\caption{Behavior of the different attenuation models from \S\ref{sec-atten} with period for thicknesses (a) 1\,m and (b) 2\,m. The models all include draft, and C contains additional (empirical) attenuation for long waves. For ease of reference in Figure~\ref{fig-bretspec}(a), wave periods of 6\,s, 7\,s and 8\,s are indicated as vertical dotted lines.
}
\label{fig-comp-atten}
\end{center}
\end{figure}

Figure~\ref{fig-comp-atten} shows the attenuation coefficients produced by the different models introduced in \S\ref{sec-atten}, computed for two different ice thicknesses. Because the C curves include an empirical inelastic contribution, they
%display the most attenuation of longer waves for smaller thicknesses as intended
{produce the greatest attenuation for large periods, and by necessity this is more pronounced for thinner ice as data are absent for larger thicknesses}.
%The B curves are all reasonably similar, although there are some slight differences between them.
Curves corresponding to model A, which are generated using random floe lengths obeying the distribution (\ref{pdf-split-power-law}) of \citet{toyota-etal2011-floedist}, are markedly different from the other curves. Due to the small values of {average floe length} $\langle D\rangle$ (in Figures~\ref{fig-comp-atten}a and b, $\langle D\rangle$ is respectively 32.5\,m and 52.6\,m), there is only a very small amount of attenuation of long waves, which qualitatively contradicts the observations of \citet{squire_moore80}, noting that their data set was for thinner ($<0.5\rm\,m$) Bering Sea ice. There is also some additional fine structure in the attenuation from model A for lower periods. In particular, there is an interval of periods between about 6\,s and 12\,s (the interval moves to higher periods as ice thickness increases), where there is much less attenuation than the other models. This will have a profound effect on the floe breaking produced by model A, as waves from that range of periods can produce very large strains if they remain unattenuated.

\begin{figure}[ht]
%\psfrag{SD, m2s}{SD, m$^{2}$s}
%\psfrag{Hs, m}{$H_{s}$, m}
%\psfrag{Es}{$E_{s}$}
%\psfrag{N}{$N$}
\begin{center}
\hspace{-31mm}\includegraphics[width=108mm]{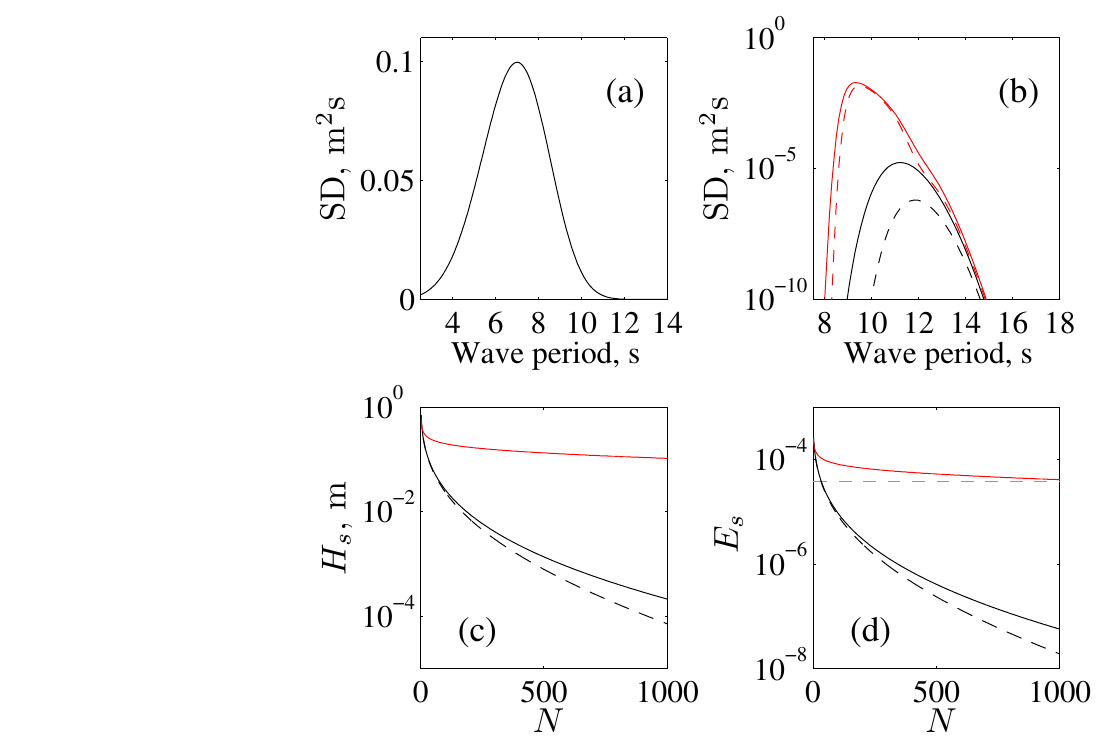}
\caption{(a): Examples of a Bretschneider spectral density when the peak period is 7\,s (green) and $H_{s}=1$\,m. (b-d): attenuation of the incident Bretschneider spectrum from (a), when the ice thickness is 2\,m. (b): spectral density (SD) after traveling past $N=250$ floes (solid curves) and $N=500$ floes (dashed curves). Two attenuation models are used; A (red) and C (black). The ice thickness is 2\,m. (c--d): Significant wave height and strain amplitude after traveling past $N$ floes, using three attenuation models: A (red), B (solid black) and C (solid black). In (d), the strain that $E_{s}$ must exceed to produce breaking,  $E_{c}$, is plotted as a dashed green line. (Here we have used $\eps_{c}=5\times10^{-5}$, $\prob_{c}=0.5$ and $N_{\rm W}=25$, so $E_{c}=3.7\times10^{-5}$.)
%{\bf %/home/nersc/timill/validation/matlab/PAPERS\_progs/NERSC/floe-breaking/miz1d:
%Bretspec.eps;
%fig\_Bretspec.m}
}
\label{fig-bretspec}
\end{center}
\end{figure}

{Figure~\ref{fig-bretspec} further illustrates the deficiencies of model A.}
%In {Figure~}\ref{fig-bretspec}(b) we take an incident wave spectrum (Bretschneider spectra with $H_{s}=1$\,m, $T_{\rm M}=7\,$s; illustrated along with spectra with $T_{\rm M}=6$\,s and 8\,s in {Figure~}\ref{fig-bretspec}a)
% and compare how they evolve under the A and C attenuation models.
{{Figures~}\ref{fig-bretspec}(a) and (b) show how a Bretschneider spectrum ($H_{s}=1$\,m, $T_{\rm M}=7\,$s) would change after being attenuated by either 250 or 500 floes,
using models A and C.}
%(This is done by letting the waves travel past 250 and 500 floes.)
The attenuated spectra produced by model A have much smaller peak periods than those of C (and B), and much larger maximum values. Figures~\ref{fig-bretspec}(c) and (d) show how both the signficant wave height $H_{s}$ and the significant strain $E_{s}$ change with $N$, the number of floes that the waves have {passed}. After only a small number of floes it can be seen that $H_{s}$ and $E_{s}$ for A (red curve) are several orders of magnitude larger than for the other attenuation models, which are roughly the same.
We can also see that for model A, $E_{s}$ remains well above the approximate breaking strain for the range of values of $N$ that are plotted. $E_{s}$ curves for B and C both drop quickly below $E_{c}$, suggesting that the width of the MIZ, $L_{\rm MIZ}$, will be similarly small under either of these models but will be significantly larger under model $A$ if strain failure is the main breakage mechanism. In fact, in our simulations involving model A, we found that our 450-km transect was almost always entirely broken. We therefore disregard model A for the remainder of the numerical results, on the basis that the predicted attenuation rates are insufficient to replicate what is observed.
%so we do not present any results involving this model.

%

\subsection{Idealized floe breaking experiments}\label{sec-ideal}
In this section we run some idealized experiments
%DD where the ice concentration and %thickness are set to be constant over the whole ice cover
and observe how the final FSD varies with the choice of WIM or attenuation model, the properties of the incident wave spectrum, and the values of the concentration and thickness. We also check the sensitivity of our results to the breaking strain and FSD parameterization, the wave speeds and whether dispersive effects can be neglected, and to numerical settings like the choice of AAS. Sensitivity to the time step and grid size is inferred from changing the wave speeds and the breaking strain.

\begin{figure}[ht]
\begin{center}
\hspace{-29mm}\includegraphics[width=108mm]{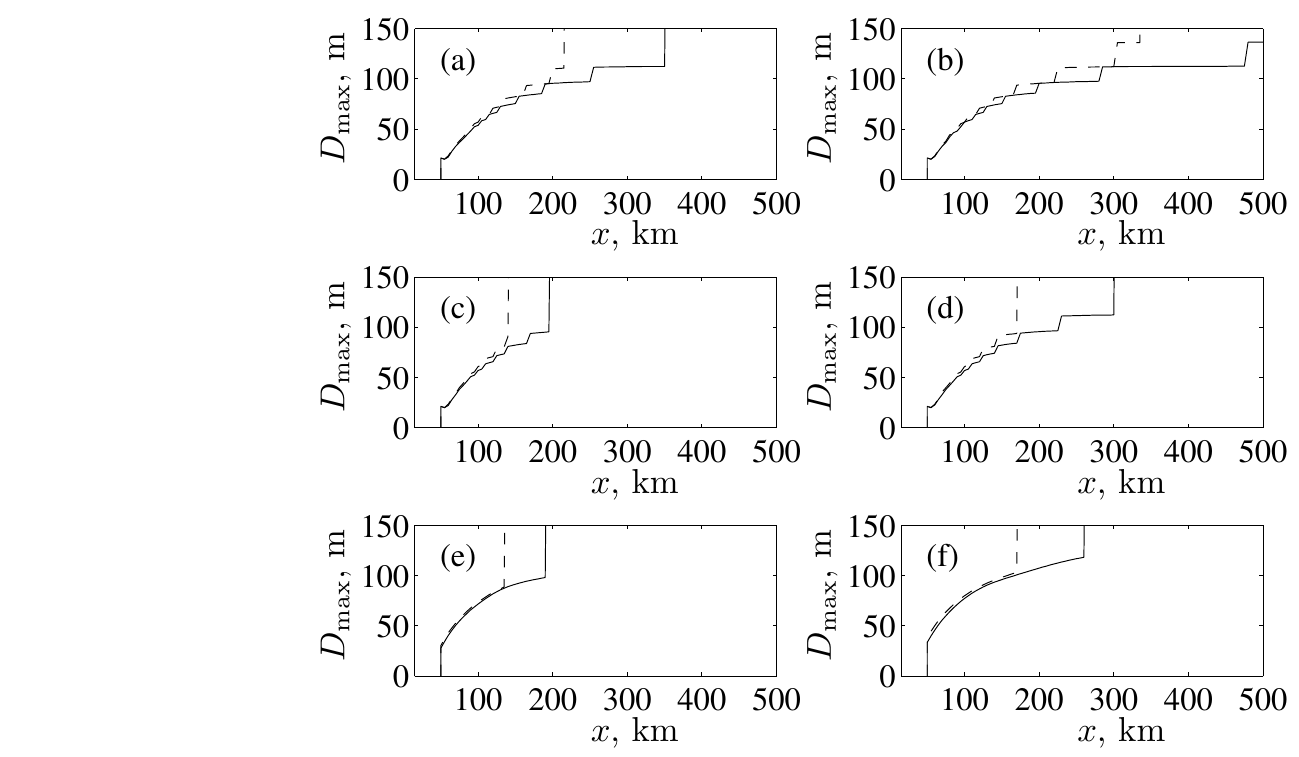}
\caption{Length distributions produced by the different attenuation models and WIMs. The WIMs used are WIM\,1 (a,\,b), WIM\,2 (c,\,d) and WIM\,3 (e,\,f). The solid curves correspond to attenuation model B, while the dashed curves correspond to attenuation model C. The significant wave height used in the Bretschneider SDF (\ref{sdf-bret}) is 3\,m, while the peak periods are 7\,s (a,\,c,\,e) and 8\,s (b,\,d,\,f). The thickness parameter $h$ is 2\,m, and the concentration has a constant value of 0.75.
}
\label{fig-lendist}
\end{center}
\end{figure}

The domain in all of our experiments is 500-km-long (consistent with our simulations in the Fram Strait in \S\ref{sec-real}), and is divided into one-hundred 5-km-wide cells. Cells 10 to 50 (counted from left to right) contain ice of constant concentration, but the thickness is given by a variable profile of the form
\begin{align}\label{thick-prof}
%h_{j}=h\left(1-e^{(x_{9}-x_{j})/x^{*}}\right)
h_{j}=
\begin{cases}
\displaystyle{h\left(1-e^{-\D x(j-9)/x^{*}}\right)} \ \ & j\geq10,\\
0 \ \ & j<10. \\
\end{cases}
\end{align}
This represents a thickness ramping up from near zero at the ice edge to a constant value $h$ after about $x^{*}=60\times10^{3}$\,m, and is included to simulate in a simple way the thicknesses at the ice edge obtained from the model data from the TOPAZ reanalysis \citep{sakov-etal2012-topaz} that is used in \S\ref{sec-real}. While not fitting Fram Strait observations precisely, the thicknesses of sea ice floes in MIZs generally do ramp up from relatively modest values near the ice edge, where the ice may be melting or being destroyed by rough seas, to larger thicknesses in the interior. Such profiles have been observed both in the Antarctic \citep{worby_etal08} and in the Greenland Sea \citep{wad83} from north of Fram Strait down to the Denmark Strait.

The left hand grid cell (cell 1) contains wave energy fitting a Bretschneider SDF (\ref{sdf-bret}) that propagates to the right into the ice, breaking it as it goes. The wave field in the left hand cell is kept constant in time, so that the final floe length distribution is in equilibrium for this sea state.

\subsubsection{Sensitivity to the choice of attenuation model and the waves-in-ice model}

Figure~\ref{fig-lendist} shows the floe length distribution as we move into the ice cover from the open water, with the different rows corresponding to when the ice is broken by WIMs\,1, 2 and 3 respectively. The two columns have different incident wave spectra; the left column has $T_{\rm M}=7$\,s and the right has $T_{\rm M}=8\,$s. Like \citet{dumont_etal2011}, we also define two parameters to summarize the floe sizes in the MIZ and the MIZ width. In the context of our results, we define the MIZ as the region of broken ice between the open ocean and the unbroken ice, where $D_{\max}$ climbs to its initial setting of 500\,m. As Figure~\ref{fig-lendist} shows, this climb is very sharp. The width of the MIZ we shall call $L_{\miz}$, and we shall label the maximum floe size in the MIZ $D_{\miz}$. For example, in Figure~\ref{fig-lendist}(c) attenuation model B produces $L_{\miz}\approx 60\,$km and $D_{\miz}\approx98$\,m.

\begin{figure}[ht]
\begin{center}
\hspace{-29mm}\includegraphics[width=108mm]{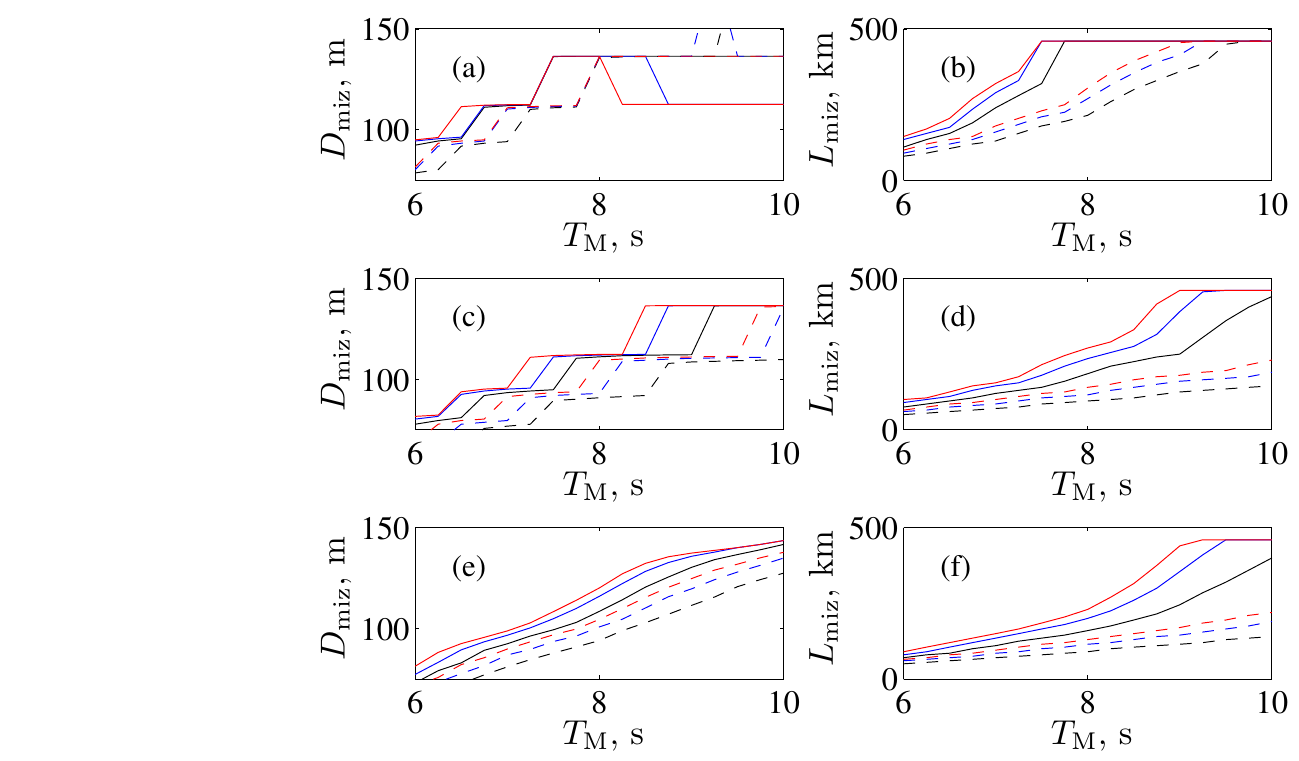}
\caption{The behavior of $D_{\miz}$ and $L_{\miz}$ with the Bretschneider peak period and significant wave height, $T_{\rm M}$ and $H_{s}$, the attenuation model and WIM used. The black, blue and red curves correspond to $H_{s}=1.5$\,m, 2.5\,m and 3.5\,m respectively, while solid and dashed curves correspond to the attenuation models B and C respectively. WIM\,1 is used in (a,\,b), WIM\,2 in (c,\,d) and WIM\,3 in (e,\,f).
The ice thickness is 2\,m and the concentration is 0.75.}
\label{fig-LDmiz}
\end{center}
\end{figure}

{Figures~\ref{fig-lendist}(a,\,b) use WIM\,1, which is the model that is most similar to that used by \citet{dumont_etal2011}. {Figures~}\ref{fig-lendist}(c,\,d) use WIM\,2, in which the stress condition has been removed. We observe that this shortens the MIZ considerably. Figures~\ref{fig-lendist}(e,\,f) use WIM\,3, in which the ISM is applied. The results are similar to those produced by WIM\,2.}

One general aesthetic difference between the results of WIMs\,2 and 3, which is also seen in Figure~\ref{fig-LDmiz}, is that the $D_{\max}$ curves are smoother using WIM\,3 since the breaking length is estimated from the entire spectrum. With WIM\,2 (and WIM\,1), the same curves {are  step-like}, since we only consider each frequency separately in the WGM method. Thus the breaking length there depends more upon the spectral discretization that we use. This smoothness of the WIM\,3 results will be numerically advantageous when trying to couple with a large-scale ice rheology. Wider MIZs could also result from using WIM\,3, caused by strains due to different frequencies interfering with one another, both constructively and destructively, potentially leading to the possibility of more breaking for a given wave spectrum. However, it seems that the opposite is actually observed, as both Figure~\ref{fig-lendist} and Figure~\ref{fig-LDmiz} show. This is possibly due to the way the wave statistics are obtained from the SDF in the WGM. As mentioned earlier, (\ref{Aexp-wgm}) overestimates the mean square amplitude for lower periods --- the ones that produce more strain and thus do the breaking in WIM\,2. Nevertheless, the difference is not significant. There are also very small differences between the values of $D_{\miz}$, which are slightly longer in WIM\,3. This is because the WGM simply picks the shortest wavelength that will break the ice, whereas the ISM considers the whole wave spectrum, and longer wavelengths that might not necessarily be able to break the ice by themselves still contribute to the integrals defining the expected value of $N_{\rm W}$ and thus pull the expected wavelength up.

When the attenuation model is changed from B to C we can see that using B generally results in wider MIZs. This is especially apparent for the higher value of the peak period $T_{\rm M}$, as might be expected since the added damping in the attenuation is designed to be stronger for longer waves.

Figure~\ref{fig-LDmiz} shows the behavior of {the MIZ's maximum floe size}, $D_{\miz}$, and its width, $L_{\miz}$, as we vary {the peak period} $T_{\rm M}$ continuously while using the different WIMs and attenuation models with some different ice thicknesses and significant wave heights. This figure shows that there is a clear difference between attenuation models B and C, with B resulting in much wider MIZs for the larger peak periods. This is particularly true for WIMs 2 and 3, and confirms the importance of including the empirically-based damping for long periods. However, this also highlights the importance of both explaining theoretically the extra attenuation that is observed, and of obtaining more attenuation measurements for large wave periods to confirm the results of \citet{squire_moore80}.

When we vary the significant wave height $H_{s}$, by looking at the transition from the black curves to blue to red, we see that, unsurprisingly, $L_{\miz}$ increases as we increase $H_{s}$. This is true for both attenuation models B and C. Similarly, $L_{\miz}$ also increases with $T_{\rm M}$, due to longer waves being attenuated less.
$D_{\miz}$ also shows a general increase with $T_{\rm M}$ (for both B and C), but it shows a lot more fine structure than $L_{\miz}$ --- particularly for WIMs\,1 and 2.
The steps in those curves are due to the spectral resolution, as the wavelengths corresponding to low periods are relatively far apart. For example, when $\om=\om_{2}$ ($T=15.82$\,s), $\lambda_\mathrm{ice}/2\approx180$\,m, but when $\om=\om_{3}$ ($T=12.36$\,s), $\lambda_\mathrm{ice}/2\approx136$\,m. In addition, in Figure~\ref{fig-LDmiz}(a), we can see in the blue attenuation model A curve (WIM\,1, $H_{s}=2.5$\,m) that $D_{\miz}$ unexpectedly decreases to 136\,m after having already increased to 180\,m. This is because the ice is completely broken and the restriction on the distance means that $\mathscr{A}(\om_{3})$ has not been able to be attenuated enough to prevent it breaking the ice in the last grid cell. If the ice cover was made long enough then $D_{\miz}$ would increase monotonically as expected.

Comparing the overall results of the different WIMs, we see that the {MIZ widths} are again much larger for WIM\,1 than the others. For smaller peak periods, $D_{\miz}$ is also larger when we use WIM\,1. Again WIMs\,2 and 3 are extremely similar, so from here on we will just assume they give approximately the same results and only compare WIMs\,1 and 3. Similarly we will assume that attenuation model C is more reliable than B but we will bear in mind that B produces wider MIZs.

\subsubsection{Sensitivity to ice thickness and concentration}\label{sec-thick-conc}
\begin{figure}[ht]
%\psfrag{Dmiz, m}{$D_{\miz}$, m}
%\psfrag{Lmiz, km}{$L_{\miz}$, km}
%\psfrag{Tm, s}{$T_{\rm M}$, s}
\begin{center}
\hspace{-28.5mm}\includegraphics[width=107mm]{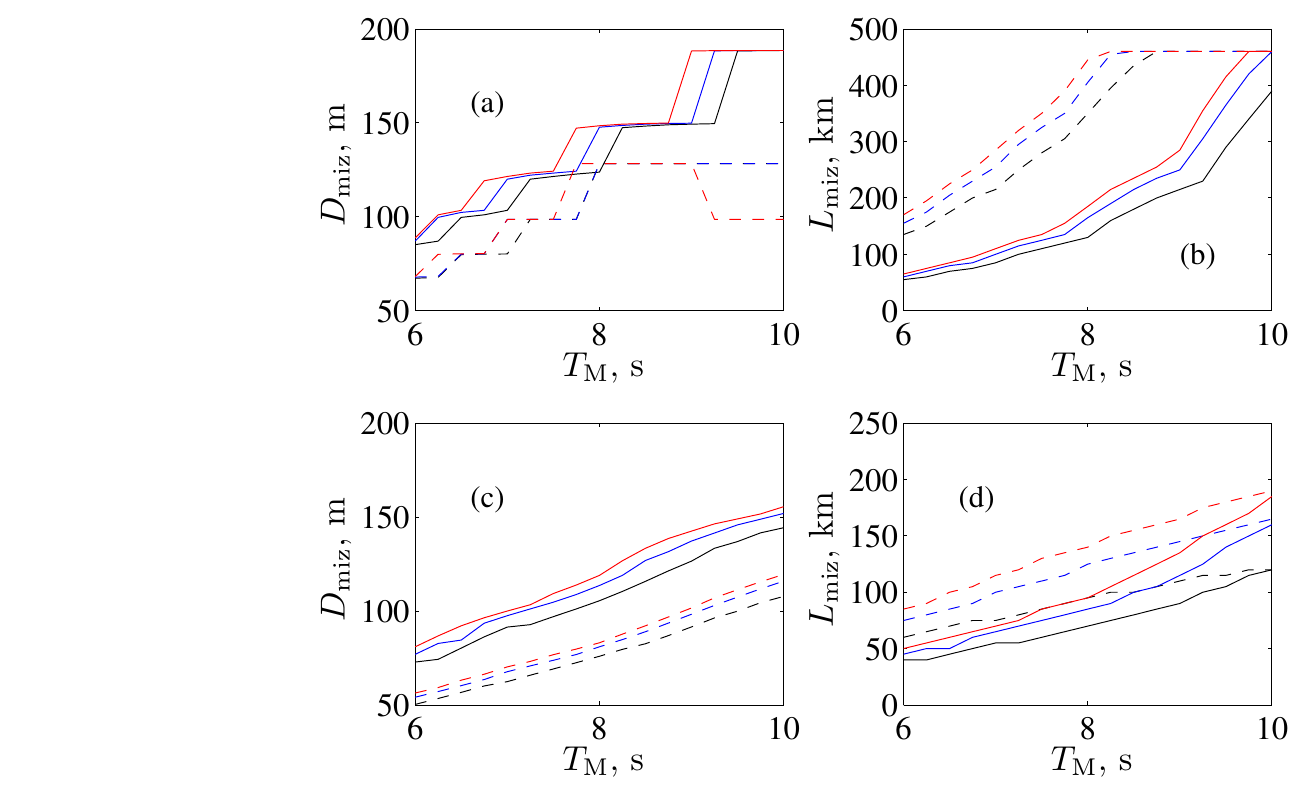}
\caption{
How $L_{\rm MIZ}$ and $D_{\rm MIZ}$ vary with ice thickness; attenuation model C with an ice concentration of 0.75 is used throughout. Black, blue and red curves respectively correspond to $H_{s}=1.5$\,m, 2.5\,m and 3.5\,m, while the dashed and solid curves designate $h=1$\,m and $h=3$\,m respectively. In (a,\,b) WIM\,1 is used while in (c,\,d) WIM\,3 is used.
%{\bf %/home/nersc/timill/validation/matlab/PAPERS\_progs/NERSC/floe-breaking/miz1d:
%compLDmiz\_thickness\_c75\_v3.eps;
%fig\_compLDmiz\_thickness\_v3.m}
}
\label{fig-thickness}
\end{center}
\end{figure}

As we increase the ice thickness we expect the MIZ width $L_{\miz}$ to decrease due to the greater attenuation associated with stiffer ice that doesn't bend so readily.
{This is indeed} observed in {Figure~}\ref{fig-thickness} when we compare the dashed curves in \ref{fig-thickness}(b,\,d) {($h=1$\,m)} with the solid ones {($h=3\,$m)}. %Note that \vern{Figure~}\ref{fig-thickness} is only plotted for model C; the corresponding plots for model A (not shown) would not give the same general result because with increased thickness the attenuation minimum moves to an interval of higher wave energy and lets more waves through that can then cause breakage.
As seen in earlier plots, {WIM\,1} produces a wider MIZ than {WIM\,3} because of the added effect of breakup induced directly by stress condition (\ref{stress-con}) in the former case. For this model, $L_{\miz}$ increases monotonically for both thicknesses with the peak wave period until it reaches 450\,km, {i.e.\ }until all the ice is broken.
The effect of significant wave height $H_{s}$ is predictable --- higher waves cause more breakup and an increase in $L_{\miz}$. The MIZ widths produced by WIM\,3 are about half the WIM\,1 widths.

Figures~\ref{fig-thickness}(a,\,c) also show that, in general, the maximum MIZ floe length $D_{\miz}$ depends strongly upon the thickness $h$, approximately doubling as $h$ increases from 1\,m to 3\,m for both models. As in Figure~\ref{fig-LDmiz}, the WIM\,1 curves exhibit a step-like structure due to the spectral resolution. However, they still generally increase within a monotonically increasing envelope. When WIM\,3 is used, in Figure~\ref{fig-thickness}(c), the curves all behave
in a simple manner, increasing smoothly and monotonically. This is also what was observed in Figure~\ref{fig-LDmiz}. The significant wave height makes little difference to $D_{\miz}$ for either model.

\begin{figure}[ht]
%\psfrag{Dmiz, m}{$D_{\miz}$, m}
%\psfrag{Lmiz, km}{$L_{\miz}$, km}
%\psfrag{Tm, s}{$T_{\rm M}$, s}
\begin{center}
\hspace{-28.5mm}\includegraphics[width=107mm]{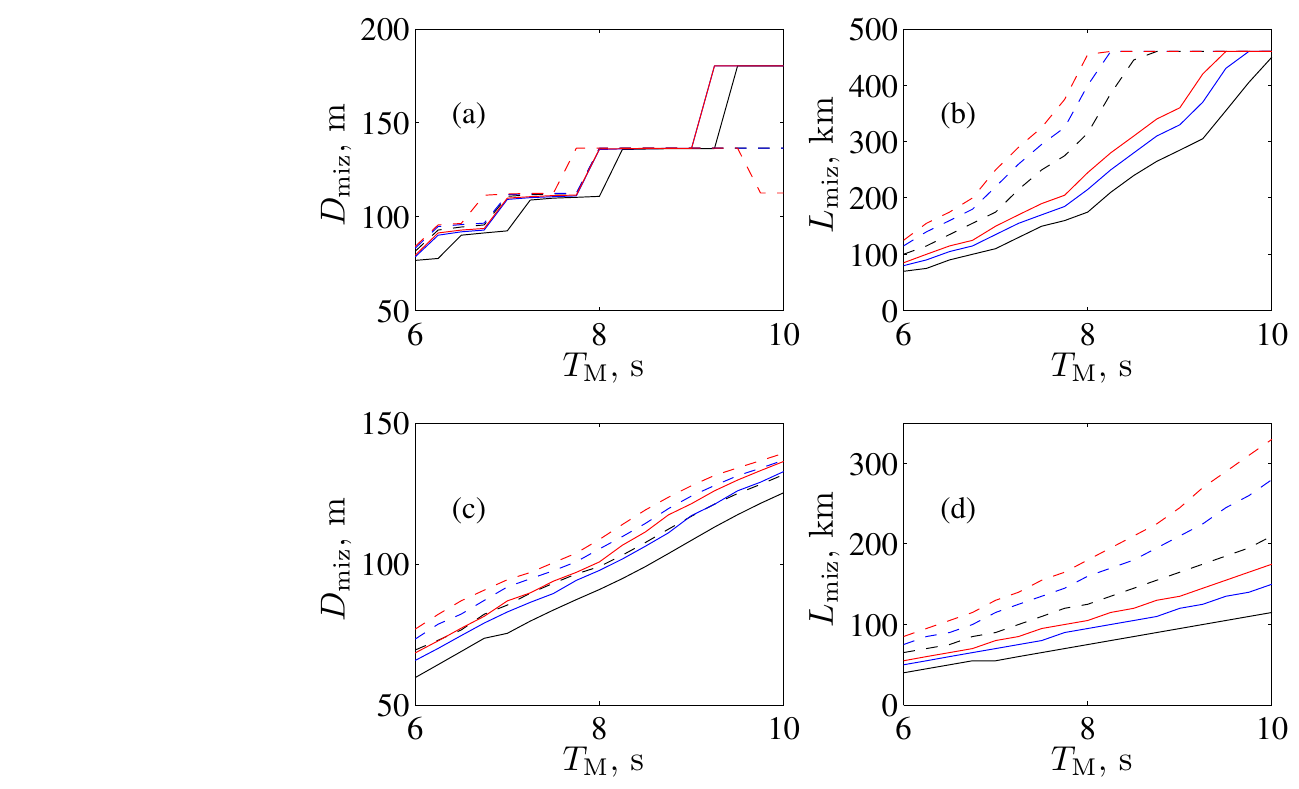}
\caption{
The effect of a change of concentration on $L_{\rm MIZ}$ and $D_{\rm MIZ}$ for attenuation model C with the ice thickness set at 2\,m. The black, blue and red curves correspond to $H_{s}=1.5$\,m, 2.5\,m and 3.5\,m respectively, while the dashed and solid curves designate $c=0.5$ and $c=0.95$ respectively. In (a,\,b) WIM\,1 is used while in (c,\,d) WIM\,3 is used.
%{\bf %/home/nersc/timill/validation/matlab/PAPERS\_progs/NERSC/floe-breaking/miz1d:
%compLDmiz\_conc\_h2\_v3.eps;
%fig\_compLDmiz\_conc\_v3.m}
}
\label{fig-conc}
\end{center}
\end{figure}

In general, an increase in concentration increases the attenuation and so causes the MIZ width $L_{\miz}$ to decrease. This is seen in {Figure~}\ref{fig-conc}, where the dashed curves correspond to $c=0.5$ and the solid ones to $c=0.95$. All curves increase monotonically with $T_{\rm M}$ and the wider MIZ that follows from a bigger $H_s$ is predictable. The $D_{\miz}$ curves in Figure~\ref{fig-conc}(c) do not appear to be affected very much by the concentration, but $L_{\miz}$ roughly doubles when the concentration is dropped to 0.5.

\subsubsection{Numerical considerations}\label{sec-adv-scheme}
There are a number of numerical approximations we can make during the implementation {of} this model. For example we would like to know {the effect of} the type of advection-attenuation scheme (AAS) we use, and the sensitivity to the time step $\D t$ and the grid size $\D x$. In addition, we will investigate whether it is admissible to approximate the wavelengths for waves in ice by the wavelengths in water, and whether it is justifiable to ignore the effects of dispersion and set the wave speed constant for all frequencies.

In results not shown, we found that approximating the wavelengths by {values for} water tended to increase the width of the MIZ only slightly, but reduced the value of $D_{\miz}$ by as much as half. Although our primary interest in the current investigation is in $L_{\miz}$, the reduced floe sizes could potentially affect other aspects of a coupled ice-ocean model (e.g.\ thermodynamics), and therefore the approximation is considered ill-advised.

However, when testing the effects of dispersion we did replace the spatially variable group velocity for waves in ice by the open water group velocity. This was because
{using} a spatially variable wave speed to allow for different ice thicknesses in different grid cells proved to be numerically problematic, primarily because of the large difference between speeds in open water and ice for small wave periods. Specifically, the CFL number of these lower-period waves $\Cop_{j,k}$ was as low as 0.03 while they were traveling in open water, and they were thus numerically attenuated almost entirely before they reached the ice. However, since low-period waves are attenuated the most strongly, we would not anticipate that approximating the group velocity in this way would have a large effect, and so the results obtained will still give us a good idea of how dispersion affects our MIZ widths and the floe sizes inside the MIZ.

\begin{figure}[ht]
%\psfrag{Dmiz, m}{$D_{\miz}$, m}
%\psfrag{Lmiz, km}{$L_{\miz}$, km}
%\psfrag{Tm, s}{$T_{\rm M}$, s}
%\includegraphics[width=80mm]{comp_advection_IL1_SS1.eps}
\begin{center}
\hspace{-28.5mm}\includegraphics[width=108mm]{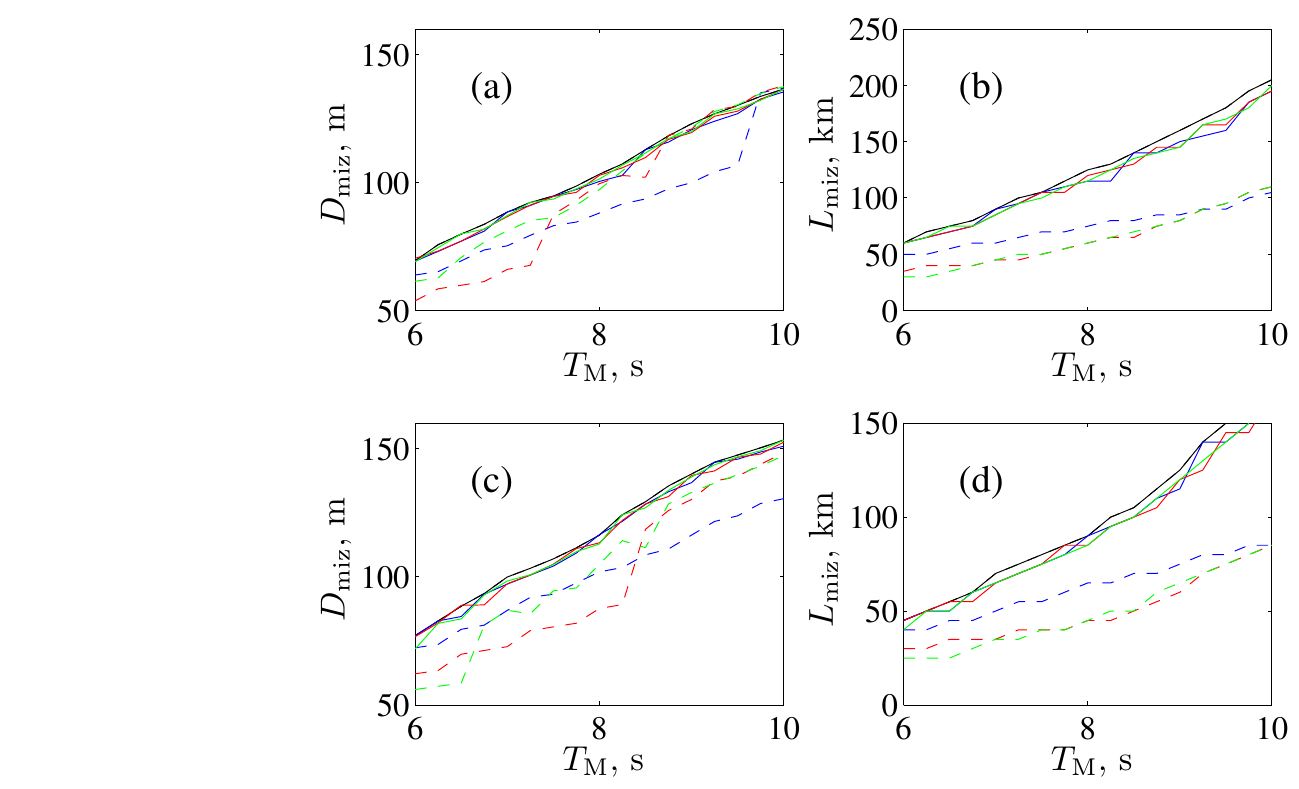}
\caption{
The effect of changing the AAS on $L_{\rm MIZ}$ and $D_{\rm MIZ}$. WIM\,3 is used with attenuation model C, the concentration is set to 0.75 and $H_{s}=3$\,m. The ice thickness is 2\,m in (a,\,b) and 3\,m in (c,\,d). The solid lines correspond to AAS\,2 and the dashed lines to AAS\,1. The time step is set to a constant 260\,s and wave speed is set to $\Cop\D x/\D t$ for all wave frequencies, where $\Cop=1$ (black curves), $0.9$ (blue curves), $0.8$ (red curves) and $0.7$ (green curves).
%{\bf %/home/nersc/timill/validation/matlab/PAPERS\_progs/NERSC/floe-breaking/miz1d:
%comp\_advection\_IL1\_SS1.eps;
%fig\_comp\_advection.m}
}
\label{fig-adv-IL1-SS1}
\end{center}
\end{figure}

In Figure~\ref{fig-adv-IL1-SS1}, we begin by looking at the advection scheme. There we set the wave speed constant for all frequencies and the time step equal to a constant $\D t=260$\,s. Physically speaking, varying the speed should thus have no effect on $D_{\miz}$ and $L_{\miz}$, and it is comforting that the results produced by {AAS\,2} show only small (numerical) variations for the different CFL numbers used, as illustrated by the close grouping of the solid curves. The solid black curve (which agrees with the dashed black {curve, i.e.\ }both schemes agree when $\Cop=1$) is the benchmark to compare the other curves against. However, the results produced by {AAS\,1} show a dramatic drop in MIZ width as the CFL number $\Cop$ is decreased. Even using $\Cop=0.9$ (dashed blue curve) consistently reduces $L_{\miz}$ by a half, and $\Cop=0.7$ (dashed green curves) reduces it by three-quarters. These results are consistent for both the ice thicknesses shown, $h=2$\,m (\ref{fig-adv-IL1-SS1}\,a,\,\ref{fig-adv-IL1-SS1}\,b) and $h=3$\,m (\ref{fig-adv-IL1-SS1}\,c,\,\ref{fig-adv-IL1-SS1}\,d). The values of $D_{\miz}$ are also smaller, but the effects are not as dramatic as they are for $L_{\miz}$. As discussed earlier in \S\ref{sec-adv}, the explanation for the difference is that in {AAS\,1}, waves are overly attenuated for $\Cop<1$, as the breaking they do is effectively moved ahead of them (since $D_{\max}$ and $\langle D\rangle$ are uniform over the whole grid cell), and they must then move through this broken ice to `escape' into the next cell. {AAS\,2} corrects for this effect by delaying {the update of the} attenuation coefficients until the waves move into the next cell.

Considering {Figure~}\ref{fig-adv-IL1-SS1} now from a physical point of view, and assuming the {AAS\,2} results are the correct ones, we see that the width of the MIZ {increases} monotonically with the peak period of the incident wave spectrum.
This is because, although the initial significant wave height is the same, the {small wave periods are the most attenuated (see Figure~2, for example)}, so $H_{s}$ decays more slowly into the MIZ as $T_{\rm M}$ increases. This generally corresponds to larger strains and more breaking further into the MIZ. However, since the dominant wave period $T_{\rm W}$ also increases with both $T_{\rm M}$ and penetration, and longer waves produce less strain, we would expect that the $L_{\miz}$ curves would reach a maximum and drop off further if $T_{\rm M}$ increases to much larger values.

\begin{figure}[ht]
\begin{center}
%\psfrag{Dmiz, m}{$D_{\miz}$, m}
%\psfrag{Lmiz, km}{$L_{\miz}$, km}
%\psfrag{Tm, s}{$T_{\rm M}$, s}
%\includegraphics[width=80mm]{comp_advection_IL1_SS0.eps}
\hspace{-27mm}\includegraphics[width=108.5mm]{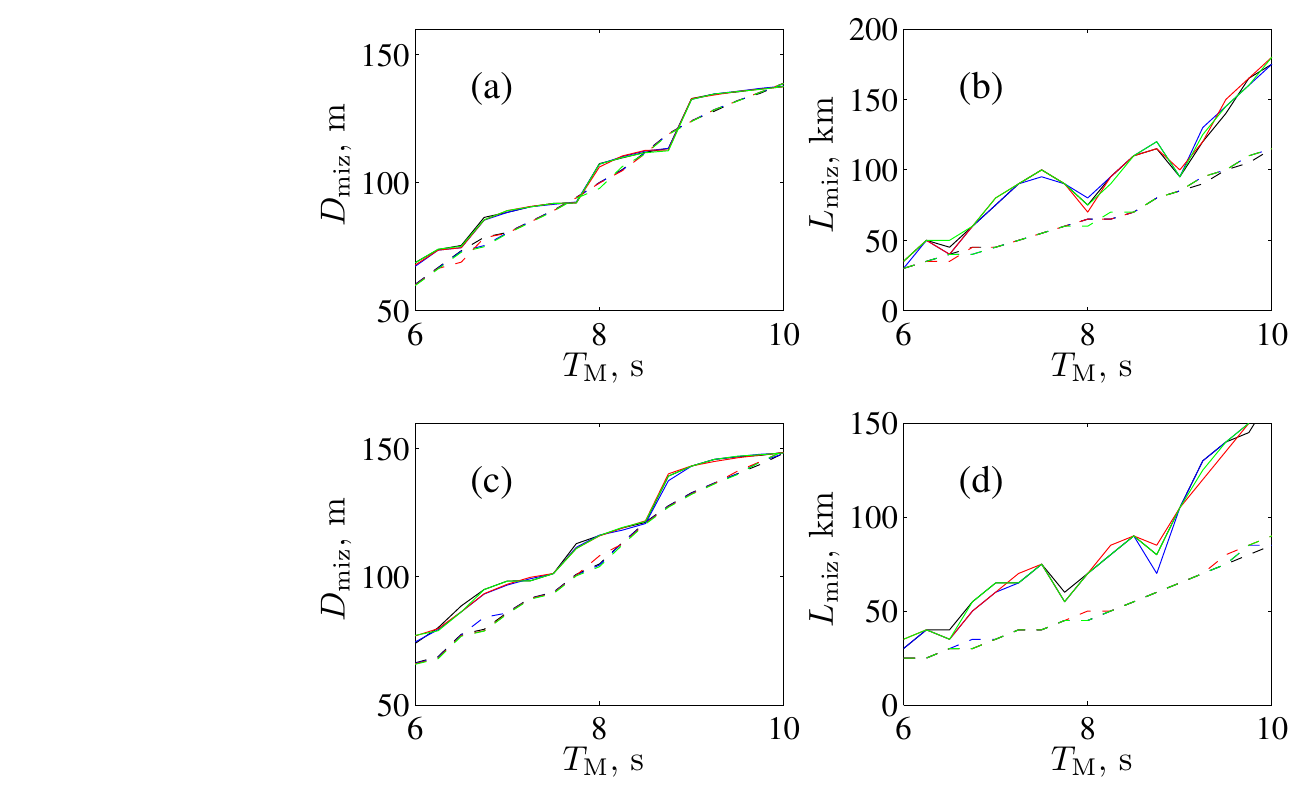}
\caption{
Same as Figure~\ref{fig-adv-IL1-SS1}, except that the effect of dispersion is accounted for, with waves of frequency $\om$ traveling with their group velocity in open water, but scaled so that the maximum speed for waves with periods between 2.5\,s and 25\,s is $(\Cop\D x/\D t)$.
%{\bf %/home/nersc/timill/validation/matlab/PAPERS\_progs/NERSC/floe-breaking/miz1d:
%comp\_advection\_IL1\_SS0.eps;
%fig\_comp\_advection.m}
}
\label{fig-adv-IL1-SS0}
\end{center}
\end{figure}

In {Figure~}\ref{fig-adv-IL1-SS0}, we observe the results produced when we introduce the effects of dispersion. In this case the results of both AASs are largely insensitive to $\Cop$, the CFL number of the fastest wave. The wave speeds are the group velocities of waves in open water, but scaled so that the fastest wave has speed $\Cop\D x/\D t$. The global time step is again set to 260\,s for consistency. The $L_{\miz}$ results from {AAS\,1} are very similar to the results shown in {Figure~}\ref{fig-adv-IL1-SS1} for the {lowest} CFL number. In general the {AAS\,2} results for $L_{\miz}$ are only slightly smaller than their {Figure~}\ref{fig-adv-IL1-SS1} counterparts, with the exception of two local minima at about $T_{\rm M}=7.5$\,s and 8.5\,s when $h=2\,$m (\ref{fig-adv-IL1-SS0}\,b) and at $T_{\rm M}\approx7.5$\,s and 8.2\,s when $h=3$\,m (\ref{fig-adv-IL1-SS0}\,d). The minima are much less pronounced for the larger thickness. Although they drop suspiciously close to the {AAS\,1} results, they are very robust with respect to changes in the CFL number and appear to be genuine physical phenomena.

To explain the presence of the minima in AAS\,2 let us reconsider {Figure~}\ref{fig-adv-IL1-SS1}. There, the leading wave-front contains waves of all frequencies traveling at the same speeds. The leading waves have only traveled through unbroken ice and are therefore relatively unattenuated. {Consequently,} a particular grid cell experiences a few time steps of large waves passing it before the more attenuated waves arrive and $H_{s}$ starts to decrease. In addition, it is during these few time steps when $H_{s}$ is at a maximum {that} breaking will happen if the waves are large {enough and, if breaking does happen, it causes additional attenuation which decreases $H_{s}$ even further.} When dispersion is introduced (as it is in {Figure~\ref{fig-adv-IL1-SS0}), each} grid cell will also experience this general pattern of $H_{s}$ building to a maximum and then dropping again as time progresses. Also, as before, the maximum $H_{s}$ and the dominant wave period $T_{\rm W}$ will increase with $T_{\rm M}$. However, the leading wave front will only consist of the longer, faster waves and the maximum significant wave height will be less than if dispersion was neglected. Moreover, the shorter waves produce more strain and they, being slower, will have also had to travel through broken ice, so less breaking will occur further in. This explains why $L_{\miz}$ is smaller with dispersion than without. Also, because $H_{s}$ is smaller in each cell the probability of breaking occurring is much closer to the critical probability. Therefore even though a slightly higher incident peak period produces higher significant wave heights further in, because $T_{\rm W}$ is also higher there may also be slightly lower strains and so the breaking probability may drop to just below $\prob_{c}$. {And,} of course, increasing $T_{\rm M}$ {further increases $H_{s}$} and so the breaking probability may increase again.

Figures~\ref{fig-adv-IL1-SS0}(a,c), and a couple of the AAS\,1 curves in Figures~\ref{fig-adv-IL1-SS1}(a,c), also show a step-like structure in $D_{\miz}$, similar to what was observed in Figures~\ref{fig-LDmiz}--\ref{fig-conc} for WIMs\,1 and 2. In that case, the steps were due to the spectral resolution, but this is not such an issue here since we are integrating the wave spectrum. The steps in Figures~\ref{fig-adv-IL1-SS0}(a,c) and Figures~\ref{fig-adv-IL1-SS1}(a,c) can better be attributed to the same mechanism that we proposed to explain the minima in the $L_{\miz}$ curves in Figures~\ref{fig-adv-IL1-SS0}(b,d).

\begin{figure}[ht]
\begin{center}
\hspace{-29mm}\includegraphics[width=108mm]{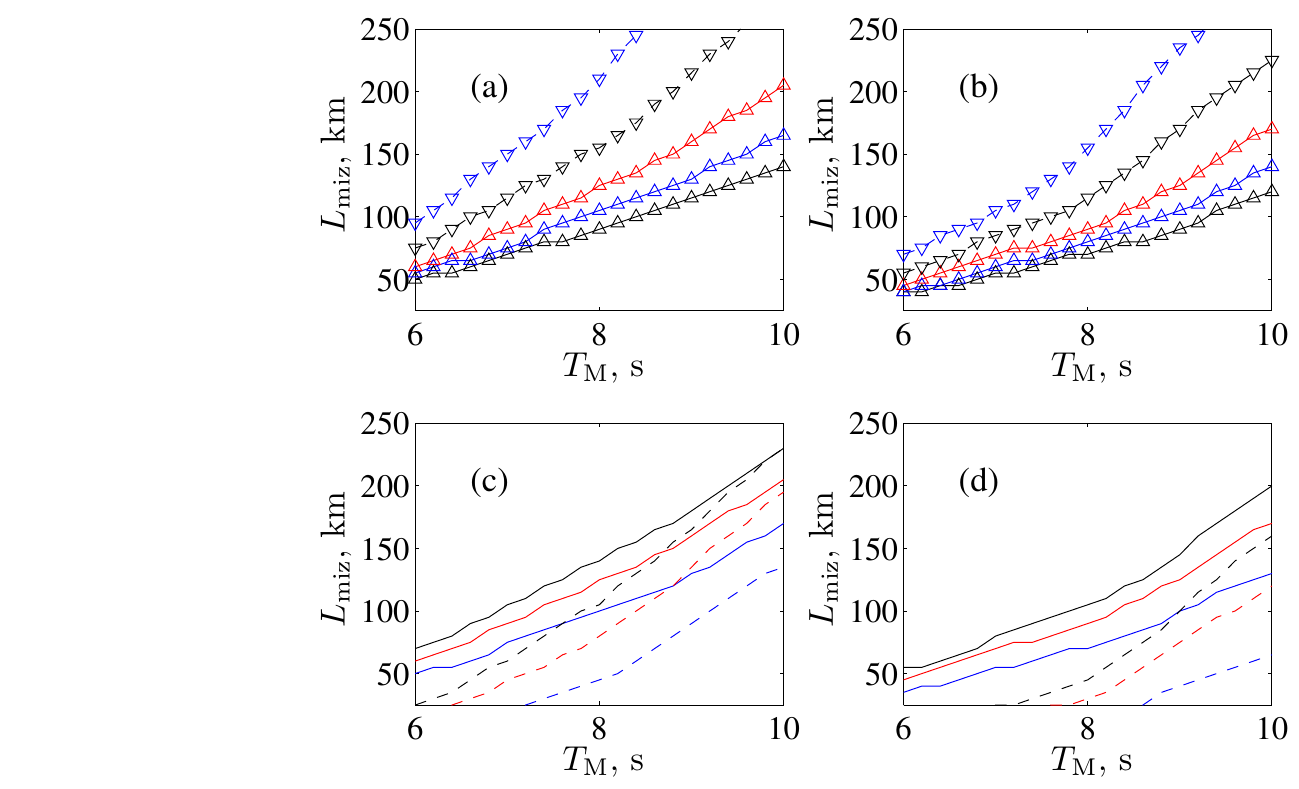}
%\psfrag{Dmiz, m}{$D_{\miz}$, m}
%\psfrag{Lmiz, km}{$L_{\miz}$, km}
%\psfrag{Tm, s}{$T_{\rm M}$, s}
%\includegraphics[width=80mm]{epsc_fsd.eps}
%\includegraphics[width=80mm]{compLDmiz_wim1_testsens_Lmiz.pdf}
\caption{
(a-b): The effect of the FSD and the value of the endurance breaking strain $\eps_{c}$ on $L_{\miz}$. In (a) $h=2$\,m, while in (b) $h=3$\,m. In both plots, when the FSD described in \S\ref{sec-fsd} is used, $L_{\miz}$ is plotted as a dashed blue curve when $\eps_{c}=10^{-5}$, a dashed black curve when $\eps_{c}=3.25\times 10^{-5}$, a red curve when $\eps_{c}=5.5\times 10^{-5}$, a solid blue curve when $\eps_{c}=7.25\times 10^{-5}$, and a solid black curve when $\eps_{c}=10^{-4}$. When the FSD is changed to the one used by \citet{dumont_etal2011}, $L_{\miz}$ is plotted for the same values of $\eps_{c}$ using the markers ${\blue\triangledown}$, $\triangledown$, ${\red\vartriangle}$, ${\blue\vartriangle}$ and $\vartriangle$ respectively. (c-d): The effect of $D_\mathrm{init}$, the initial value of $D_{\max}$, and the shape of the thickness profile on $L_{\miz}$. In (c) $h=2$\,m, while in (d) $h=3$\,m. Blue, red and black curves correspond to $D_\mathrm{init}=250$\,m, 500\,m and 750\,m respectively. Solid curves correspond to thickness profiles of the form (\ref{thick-prof}), while dashed curves correspond to a uniform thickness being used. In all four plots, %(a-d),
WIM\,3 is used with attenuation model C and $H_{s}=3$\,m.
%{\bf %/home/nersc/timill/validation/matlab/PAPERS\_progs/NERSC/floe-breaking/miz1d:
%espc\_fsd.eps;
%fig\_espc\_fsd.m}
}
\label{fig-testsens}
\end{center}
\end{figure}

\subsubsection{Other sensitivities}\label{sec-testsens}
In Figures~\ref{fig-testsens}(a,\,b) we investigate the sensitivity of the MIZ width to the FSD parameters, and the parameters involved in the breaking criterion (\ref{strain-con}) for WIM\,3: the breaking strain $\eps_{c}=\sig_{c}/(Y^*(1-\nu^{2}))$, the number of waves per time step $N_{\rm W}$ and the probability threshold $\prob_{c}$.
Since $E_{c}$ is linear in $\eps_{c}$, we expect that it will produce the most variability of the latter three. For example, if $N_{\rm W}$ is taken from the normal distribution $\Nop(25,5)$ and $\prob_{c}$ from $\Nop(0.5,0.15)$, where $\Nop(m,s)$ has mean $m$ and standard deviation $s$, then varying these two parameters only produces a standard deviation in $E_{c}$ of about five or six percent. Also, we note that varying $N_{\rm W}$ acts as a proxy for varying the time step. Therefore, if we let $\eps_{c}$ experience a 10--15\% range of variability we should be able to infer the effects of changing $\D t$ and $\prob_{c}$. However, although the value of Poisson's ratio $\nu$ appears to be reasonably consistent for sea ice \citep{langleben-pounder1963}, the value we use for $Y^*$ and hence $\eps_{c}$ contains a large degree of uncertainty. As a result, we have let $\eps_{c}$ vary over an order of magnitude to obtain a more conservative estimate of its effects.

The curves corresponding to different values of $\eps_{c}$ in Figure~\ref{fig-testsens}, show that $L_{\miz}$ is highly sensitive to this parameter. This is particularly apparent in Figure~\ref{fig-testsens}(a) when $h=2$\,m. The middle curves in both figures relate to $\eps_{c}=5.5\times10^{-5}$, the value resulting from using $\upsilon_{b}=0.1$ in (\ref{sig-c-formula}--\ref{eps-c-formula}). Clearly reducing $\eps_{c}$ makes more difference than increasing it, but there is still considerable variation for the higher values. Nevertheless, $7.25\times10^{-5}$ is probably close to the upper limit for $\eps_{c}$, while $4\times10^{-5}$ is close to the lower limit, so our MIZ widths have an uncertainty of about 25\% in them due to uncertainties in $\eps_{c}$. Likewise, the variability caused by $\D t$ and $\prob_{c}$ is about half this.

\begin{figure}[ht]
\begin{center}\hspace{-4mm}
\includegraphics[width=80mm]{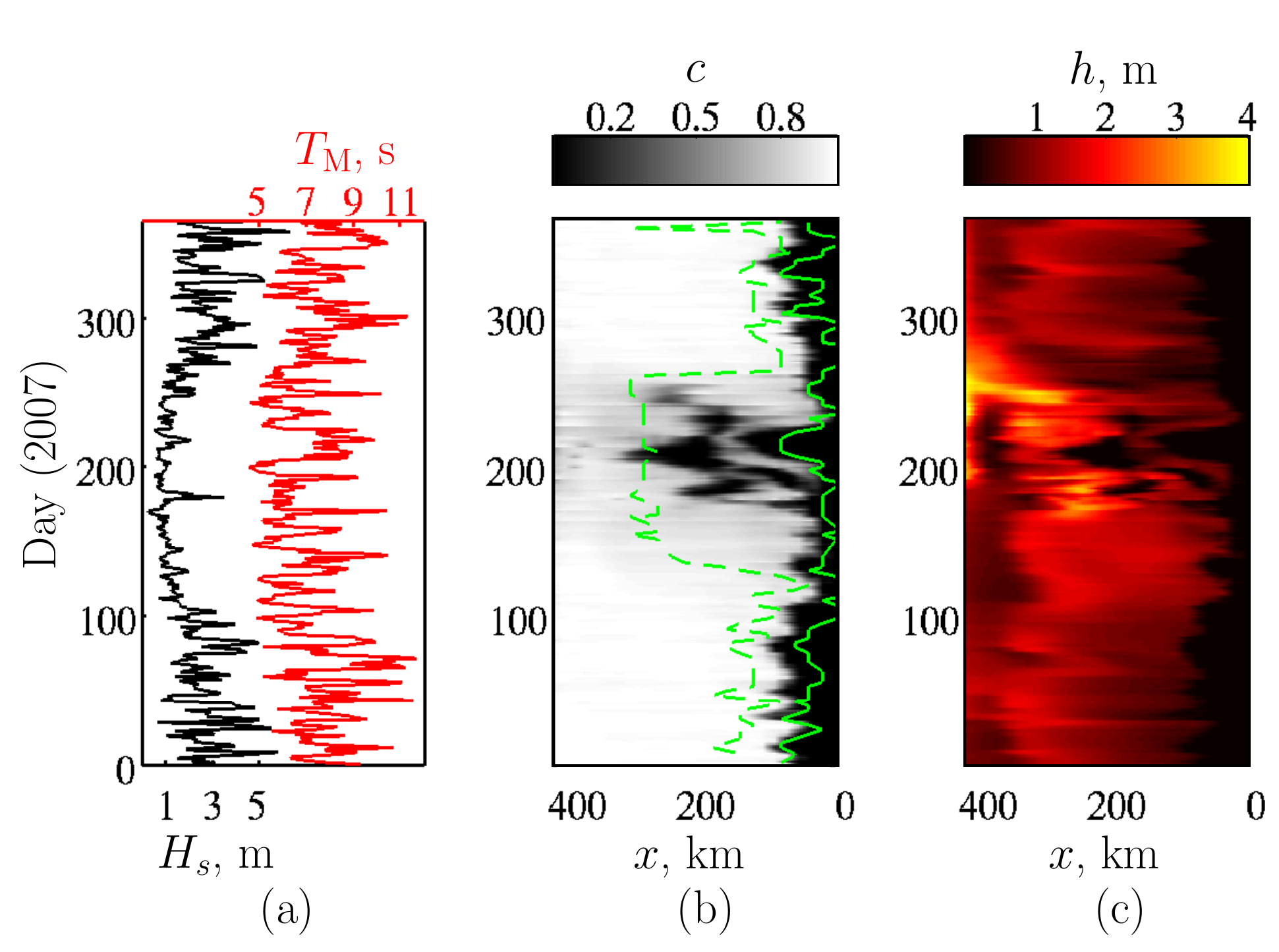}
\caption{
Model data for our one-dimensional simulations in the Fram Strait in 2007,
between the south-east coast of Norske {\O}er (latitude 79$^{\circ}$N, longitude 17.7$^{\circ}$W), which corresponds to $x=438$\,km on our one-dimensional grid, and latitude 79$^{\circ}$N, longitude $3^{\circ}$E, which corresponds to $x=0$\,km. The wave field is specified at $x=0$\,km and is obtained from the WAM ERA-Interim reanalysis. The significant wave heights and peak periods are plotted in {(a)}. The waves are then advected west through ice with concentrations and thicknesses taken from a TOPAZ reanalysis. They are interpolated onto a regular grid with longitudinal resolution of 0.125$^{\circ}$ ($\D x=2.65$\,km), and are plotted in {(b)} and (c). For comparison, the ice edge and edge of the MIZ estimated from AMSR-E concentrations are {also} plotted in (b) as solid and dashed green lines respectively.
}
\label{fig-topaz-in}
\end{center}
\end{figure}

Comparing the two FSDs used in Figures~\ref{fig-testsens}(a,\,b), i.e.\ the curves and their corresponding markers, we can immediately see that only negligible differences in the large-scale properties are produced by changing the FSD from the one we introduced in \S\ref{sec-fsd} to the one used by \citet{dumont_etal2011}.
Of course, the differences would be visible on a smaller scale. It is worth noting that we have treated all floes as experiencing the same stress field. In actual fact, the strains in a floe are dependent on its length and shape and accounting for this could increase sensitivity to the FSD. However, determining where and how a two-dimensional floe would break in such a study is not trivial and will not be attempted here.

Moving on to Figures~\ref{fig-testsens}(c,\,d) we observe that the initial FSD has more influence. When the thickness profile from (\ref{thick-prof}) is used (solid curves), we see that varying the initial value of $D_{\max}$, which we shall refer to as $D_\mathrm{init}$, between 250\,m and 750\,m causes a variation in $L_{\miz}$ of about 10--25\%. 
There is more variation (25--45\%) when we use a constant thickness profile, and we note that the 
MIZ widths are more narrow than their variable thickness counterparts.
This is because, for the variable profile, waves travel with little attenuation through the thin ice near the edge, and the ice is therefore broken.
Further in, the thicker ice causes rapid attenuation, thus acting like a barrier, and preventing further breakage.
%For the constant profile the attenuation rate does not vary.
%Consequently, it results in narrower  MIZs as the ice at the edge produces more attenuation than the thinner ice there using profile (\ref{thick-prof}). 
In \S\ref{sec-real}, when the variation in the ice thickness is increased, we shall find that this reduces the sensitivity to $D_\mathrm{init}$ to around 10\% for most of the year.

\subsection{Realistic experiments in the Fram Strait}\label{sec-real}
In this section we show the results of some simulations using WIM\,3 with realistic wave forcings, ice concentrations and  ice thicknesses for the Fram Strait during 2007.
Figure~\ref{fig-topaz-in}(a) shows the wave forcing, which {was} obtained from the ERA-Interim reanalysis (WAM model), while Figures~\ref{fig-topaz-in}(b,\,c) show ice concentrations and thicknesses obtained from a TOPAZ reanalysis \citep{sakov-etal2012-topaz} in which concentration data derived from AMSR-E  (University of Bremen) has been assimilated. On average, the modeled ice edge is 45\,km west of the ice edge observed by AMSR-E, which is plotted as a solid line in \ref{fig-topaz-in}(b). This is well within the uncertainties and resolution of the model (TOPAZ has a resolution of about 13\,km) and the resolution of the AMSR-E analysis (the transect from 15$^{\circ}$W to 5$^{\circ}$E was divided into bins with widths of about 21.2\,km, i.e.\ 1 degree in longitude, and analyzed for ice concentration \citep{kloster-sandven2011-Fram79N-flux}). The internal concentrations also compare well.

\begin{figure}[ht]
\begin{center}
\hspace{-5mm}
\includegraphics[width=82mm]{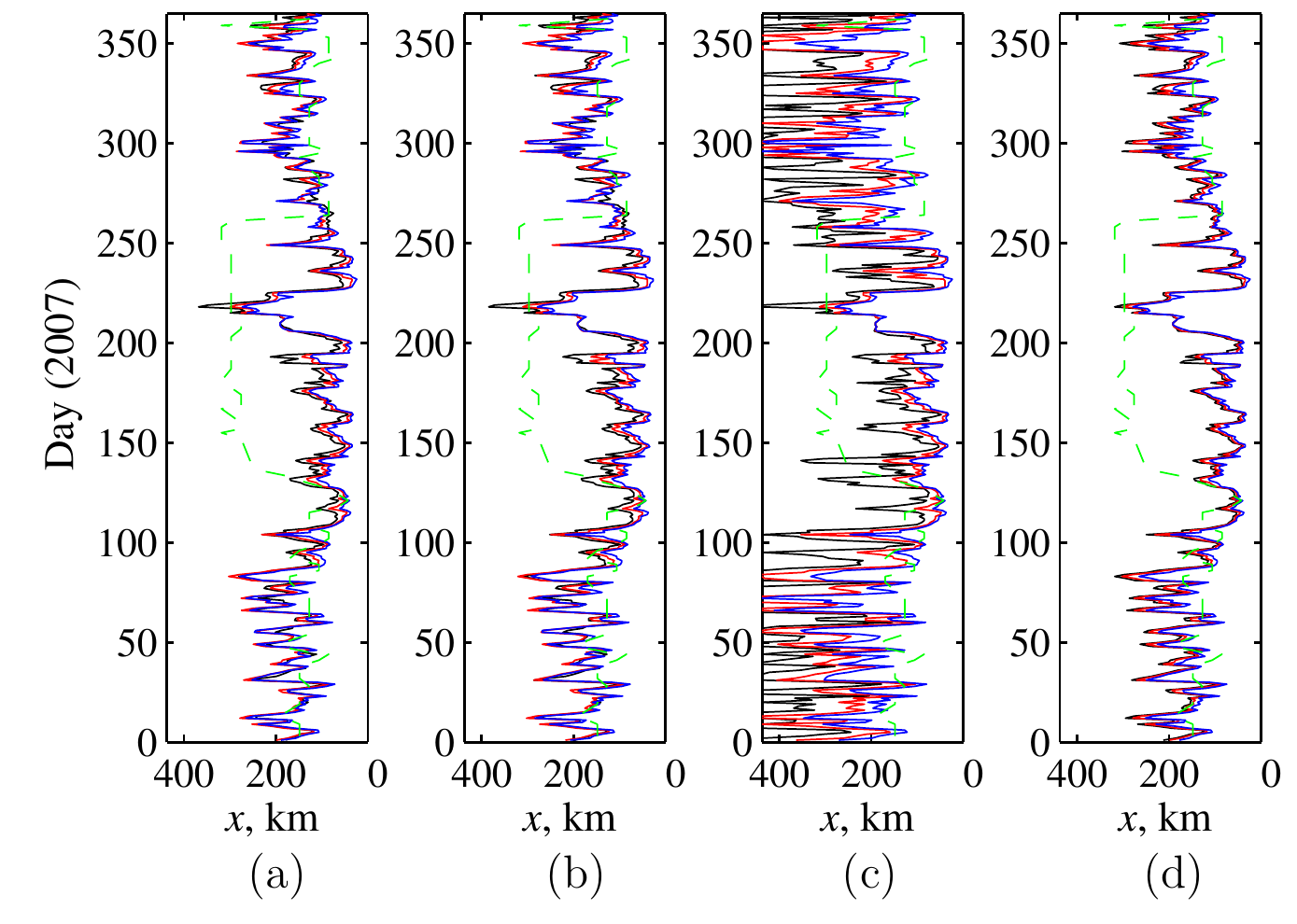}
\caption{
Results of one-dimensional simulations in the Fram Strait in 2007.
(a): the lines separating broken and unbroken ice, as determined by {WIM\,3} using attenuation model C, are plotted for $\b_{h}=1$ (black line), $\b_{h}=1.75$ (red line) and $\b_{h}=2.5$ {(blue line)}, where $\b_{h}$ is a factor used to increase the ice thicknesses from Figure~\ref{fig-topaz-in}(c) which are unrealistically low. The breaking strain $\eps_{c}$ is $5.5\times10^{-5}$, determined from (\ref{eps-c-formula}) using $\upsilon_{b}=0.1$, and the initial value of $D_{\max}$ in each grid cell is $D_\mathrm{init}=500$\,m. (b): same as (a) but with $\eps_{c}$ reduced by 25\% to $4.125\times10^{-5}$. (c): same as (a) but using attenuation model B. (d): same as the red curve in (a), but using $D_\mathrm{init}=250$\,m (black curve), 500\,m (red curve) and 750\,m (blue curve).
For comparison, the edge of the MIZ estimated from AMSR-E concentrations is plotted as a dashed green line in all plots.
}
\label{fig-topaz}
\end{center}
\end{figure}

Also plotted (dashed line) in Figure~\ref{fig-topaz-in}(b) is an estimate for the inner edge of the MIZ, determined from the same AMSR-E concentrations using the criterion that $c<0.9$ corresponds to the MIZ. While this is a different criterion from the floe size criterion we use in this paper, it should still give us a rough idea of how well our WIMs are performing.

The mean ice thickness is roughly 0.8\,m, creeping up towards 2\,m in the summer, which is thicker due to greater movement south of multi-year ice from the Arctic Ocean at that time. These ice thicknesses are probably too low, so we have also run simulations in which the ice thicknesses are multiplied by a factor $\b_{h}$ of either 1.75 or 2.5 in order to get closer to observations \cite[e.g.][]{widell-etal2003-drift-thickness}.

Figure~\ref{fig-topaz} shows the results of numerical experiments using the model outlined in this work. Results for the expected breaking are calculated using either the TOPAZ thicknesses or the increased thicknesses, the TOPAZ concentrations and the ERA-Interim waves. The values of $\eps_{c}$ used in each are $5.5\times10^{-5}$ in \ref{fig-topaz}(a,\,c\,d) and $4.125\times10^{-5}$ in \ref{fig-topaz}(b), and the attenuation model used in \ref{fig-topaz}(a,\,b\,d) is either C or B in \ref{fig-topaz}(c). Each night in \ref{fig-topaz}(a-c), the floe sizes are re-initialized to being uniformly $D_\mathrm{init}$=500\,m long. An extension to including a memory in each cell of $D_{\max}$ and a gradual refreezing was rejected on the basis that we are unable to include the more important effects of ice movement due to winds and current in this one-dimensional experiment.

In Figure~\ref{fig-topaz}(d), the value of $D_\mathrm{init}$ is varied to test the model's sensitivity to this parameter. With the inclusion of the more realistic variations in ice thickness, this is usually less than 10\%.

In all plots, except for Figure~\ref{fig-topaz}(c) when the attenuation model B is used, the edge of the broken ice compares well with the AMSR-E-determined MIZ edge in the non-summer months especially,
recalling that the ice edge from TOPAZ is on average about 45\,km too far to the west.
In the summer,  {less breaking occurs} due to smaller waves and thicker ice. It is also more likely that, due to the ice being more dilute in the summer, a concentration criterion for the MIZ will disagree with a floe size criterion. The results presented here show that our simulations have a very low sensitivity to the breaking strain, and generally do not vary much with thickness, which is one of the least well-known properties. The variations that do occur are systematic in that increasing the thickness makes the MIZ narrower, as observed in \S\ref{sec-thick-conc}. The variations with respect to $\eps_{c}$ are consistent with our previous findings, with smaller values resulting in a  wider MIZ.
As with the sensitivity tests done on $D_\mathrm{init}$, there is less variation than was observed in the idealized tests in \S\ref{sec-testsens}.

However, Figure~\ref{fig-topaz}(c) shows that using attenuation model B produces much more variation with thickness, with far too much breaking produced for the lower thicknesses. This re-emphasizes the importance of obtaining a better understanding of the attenuation process, and in particular the lower-than-observed attenuation of long waves that arises due to inelastic processes. It also highlights the urgent need for more measurements of attenuation, and also of more reliable thickness data.

\section{Conclusions}

There are many observations that suggest a primary role for ocean waves in shaping the morphology of ice fields, especially in the vicinity of the ice edge where waves are particularly fierce, but also throughout the MIZ where they habitually limit the size of the constituent ice floes by fracturing those floes that are too large to exist as the waves permeate further into the ice pack. Attenuation, arising due to scattering and supplementary inelastic processes such as turbulence, bending hysteresis and interfloe collisions and rafting, also occurs causing a gradual reduction of the wave energy envelope with distance from the ice edge that, \emph{c{\ae}teris paribus}, results in a gradual increase in floe size with penetration. The FSD is therefore continuously modified by pervasive incident ocean wave trains that, according to their period, may either travel long fetches from distant storms or else be more locally generated. They are then preferentially filtered by the sea ice in a manner that favors the survival of longer wavelengths. Notwithstanding these outcomes, waves can also redistribute ice floes spatially by herding them together, potentially creating zones of higher or lower concentration, and also polynyas, and they can assist complementary thermodynamic processes to melt the ice cover.

Given these several influential factors relating to the composition of MIZs, in particular, it is perhaps surprising that wave-ice interactions are not yet included in ice/ocean models and oceanic general circulation models. While this has been discussed in the past --- indeed VAS proposed it in the early eighties --- the complexity of doing it has proved insuperable until now. The current paper, which provides a road map of how to do it, gives some first predictions of how floe size and MIZ width are manipulated by waves in a one-dimensional spectral setting, and a realistic example of the adapted HYbrid Coordinate Ocean Model ice/ocean code in action. It continues the progress made by \citet{dumont_etal2011} towards embedding a ubiquitous and dominant physical process in forecasting ice/ocean models that are used for operational as well as research purposes. It sets the scene and provides the machinery to deal with the next stage of development, which is to incorporate two-dimensional interactions arising from a directional sea comprising energy at a comb of different frequencies distributed angularly and provided either by observations or a wave model such as WAM or WAVEWATCH III\textsuperscript{\textregistered}.

We summarize here the most substantive observations and conclusions we have reached during the reported research programme.
\begin{enumerate}
\item Three models describing how waves break up the sea ice were tested; WIMs\,1 and 2 use individual wave groups, while WIM\,3 allows wave frequencies to interact. WIMs\,2 and 3 break ice when the strain $\varepsilon_c$ is exceeded, while WIM\,1 also applies a stress condition for direct comparison with \citet{dumont_etal2011}. WIMs\,2 and 3 produce very similar results, while WIM\,1 generally produces MIZs that are too wide. This is because long waves in WIM\,1, which are relatively unattenuated, are (unphysically) more likely to produce breaking than short waves.
\item Various options exist to describe the way ocean waves attenuate as they propagate through open ice fields, differing in the type of floe size distribution used, the averaging process employed and any imposed approximation to facilitate solution. Three models are used in this paper; they are labeled A, B and C. Models~B and C are anchored in the recent work of \citet{bennetts-squire2011-atten} who recognized
that accurate information about wave phases could not be provided by two-dimensional scattering models (one horizontal and one vertical dimension), and so phases should simply be assumed to be uniformly distributed. Although it was derived from a supposedly realistic FSD, model~A did not do this with the result that artificial resonances were created that produced the over-transmission of high-curvature waves and thus too much breaking. Model~C also included an empirical parameterization of wave energy loss from inelastic phenomena other than scattering. This additional damping made a large difference to our results, giving MIZ widths that were roughly half those produced by model~B. This highlights the importance of resolving the problem of how long waves are attenuated theoretically, and also of conducting more experiments to confirm the observations of  \citet{squire_moore80} and to extend them to different thickness ranges.
\item Various sensitivity tests were done using  the different WIMs. It was found that a larger significant wave height gives a wider MIZ, which is common sense, and that the higher the ice concentration the narrower is the MIZ and the smaller is the maximum floe size. Increasing the ice thickness also decreased $L_{\miz}$, but increased $D_{\rm MIZ}$. MIZ {width} increased monotonically with the peak period of the incident spectrum.
\item It was found that different parameterizations of the FSD had only a negligible effect on MIZ width and maximum floe size. The initial FSD had a noticeable influence on the idealized results of \S\ref{sec-ideal} but less influence in the realistic simulations of \S\ref{sec-real}.
\item The results in \S\ref{sec-ideal} were also found to depend strongly on the failure strain $\varepsilon_c$ of sea ice. While very few measurements of $\varepsilon_c$ are available, a large data set has been compiled for flexural strength that is consistent with $\varepsilon_c$ at the strain rates concerned.
Nevertheless it still contains large uncertainties and should be investigated further experimentally. Again, however, the realistic simulations showed less sensitivity to this parameter.
\item Two advection and attenuation schemes (AASs) are used as described in \S\ref{sec-adv}, AAS\,2 being more physically realistic. The less sophisticated AAS\,1 was found to produce too great a dependence of MIZ width on wave speed, while AAS\,2 performed well.
\item The realistic simulations in Fram Strait, 2007 compare reassuringly well with contemporaneous AMSR-E (University of Bremen) data. They were conducted using the WAM ERA-Interim reanalysis to provide the wave forcing and a TOPAZ reanalysis for ice concentrations and thicknesses.
\end{enumerate}

\appendix

\section{Elastic beam and plate results}\label{app-EB}
Ice flexural strength tests are typically analysed using Euler-Bernoulli beam theory (EBBT) \citep{meirovitch2010-EBref}.
%{\bf**Possibly should be careful with this as probably not linear near breaking point**}
Suppose the neutral axis of an unloaded Euler-Bernoulli beam runs along the $x$ ($x_{1}$) axis, and suppose the $z$ ($x_{3}$) axis points upwards, in the opposite direction to the loading direction. Also let the beam have  width $b$, thickness $h$ and length $L$. If $z\in[-h/2,h/2]$ and $x\in[-L/2,L/2]$, the EBBT approximates the strain $\eps_{11}$ as $\eps_{11}=z\pa_{x}^{2}w$, where $w(x)$ is the vertical deflection of the beam, and relates the stress $\sig_{11}$ to $\eps_{11}$ by the constitutive relation $\sig_{11}=Y\eps_{11}$, where $Y$ is the Young's (elastic) modulus. The other property of the beam that we need to apply EBBT is the second moment of area, given by
\begin{equation}
I=\int_{-b/2}^{b/2}\int_{-h/2}^{h/2}z^{2}\rmd z\rmd y=\frac{bh^{3}}{12}.
\end{equation}
The equation of motion for an Euler-Bernoulli beam equation is
\begin{equation}\label{EBbeam0}
\pa_{x}^{2}\left(YI\pa^{2}_{x}w(x,t)\right)=q(x,t)-\rho_\textrm{ice} bh\pa_{t}^{2}w(x,t),
\end{equation}
where $q$ is the load per unit length and $\rho_\textrm{ice}$ is the ice density. (For a static formulation, such as during a flexural strength measurement, $q(x,t)=q(x)$ and $w(x,t)=w(x)$, so any time derivatives of $w$ disappear.) If $Y$ and $I$ are constant, then (\ref{EBbeam0}) becomes
\begin{equation}\label{EBbeam1}
\frac{Yh^{3}}{12}\pa_{x}^{4}w(x,t)=\frac{q(x,t)}b-\rho_{i} h\pa_{t}^{2}w(x,t),
\end{equation}
where the forcing $q/b$ is now a pressure.

In Appendix~\ref{app-3pt} we solve this equation in one of the three most common testing situations to find the stresses and strains in the beam and thus also the breaking strain and flexural strength. (We do this for the three-point-loaded beam test). The expression for the flexural strength (\ref{sigc-3pt}) derived in this way is found to be the same as given by \citet{schwarz-etal1981-icestrength}, and the associated breaking strains allow us to use the wealth of flexural strength measurements to inform our strain breaking criteria. This is not a new result, but we think it is helpful to have a derivation written down in one easily accessible place as \citet{schwarz-etal1981-icestrength} neither derive their expressions nor give any references which explain them.

After this we present formulae for the stresses and strain in an Euler-Bernoulli thin elastic plate, which is a more accurate model for an ice floe in the MIZ than the beam model. The essential difference between the two models is that for a beam, the width $b$ is small enough that Poisson's effect is not important --- the top of the beam is under compression, so the width there increases slightly, while the bottom of the beam is in tension so the width shrinks --- thus on average the width stays roughly the same. For a plate, however, the width is of the same order as the length, so this cancellation is not present to the same extent, and strains in the $y$ direction must be accounted for. This effectively makes a plate stiffer than a beam, as to bend it in one direction you must bend it in the perpendicular direction as well. This can also be seen by comparing (\ref{EBbeam1}) to the Euler-Bernoulli thin elastic plate equation
\begin{equation}\label{EBplate}
\b_\textrm{ice}\nabla^{4}w(x,y,t)=p(x,y,t)-\rho_\textrm{ice} h\pa_{t}^{2}w(x,y,t),
\end{equation}
where $p(x,y,t)$ is an applied pressure, $\nabla^{4}=(\nabla^{2})^{2}$, $\nabla^{2}=\pa_x^{2}+\pa_y^{2}$, and
%\[\b_\textrm{ice}=\frac{Yh^{3}}{12(1-\nu^{2})}\]
$\b_\textrm{ice}=(Yh^{3}/12)/(1-\nu^{2})$
is the flexural rigidity \citep{fung65}. The quantity $\nu<1$ therein is Poisson's ratio and the $1-\nu^{2}$ term increases the stiffness of the plate by roughly 10\% (for $\nu\approx0.3$) from what it was in the beam equation ($Yh^{3}/12$).

An important consequence of this is that the same stresses produce smaller strains in a plate than in a beam, making a plate harder to break. In other words, even though the stress may exceed the flexural strength for a beam, the strain may not exceed the breaking strain (which we assume is the same for an ice beam and an ice plate) and the plate will not break.

%\subsection{Cantilevered beam}\label{app-cant}
%
%To perform an {\it in situ} cantilever test, ice is cut away from an edge to leave a rectangular beam of length $L$, width $b$ and thickness $h$, joined at one end ($x=-L/2$) to the original ice block. A downward force $F$ is then applied to the free end ($x=L/2$) until the ice breaks at the root. (When this happens $F=F_{c}$.) Then the displacement of the beam from its equlibrium position, $w$, must satisfy
%\begin{subequations}
%\begin{align}
%YI\pa_{x}^{4}w(x)&=0,\\
%w(-L/2)&=0,\\
%\pa_{x}w(-L/2)&=0,\\
%YI\pa_{x}^{2}w(L/2)&=0,\\
%YI\pa_{x}^{3}w(L/2)&=F.
%\end{align}
%\end{subequations}
%The solution to this is problem is
%\[w(x)=A(x+L/2)^{2}+B(x+L/2)^{3},\ \ A=-3BL=-\frac{FL}{2YI}.\]
%The strain in the ice is thus
%\[\eps(x)=z\pa^{2}_{x}w(x)=2zA+6zB(x+L/2),\]
%which has a maximum absolute value of
%\begin{equation}\label{max-strain-cant}
%\eps_{\max}=h|A|=\frac{FhL}{2YI}
%\end{equation}
%when $x=-L/2$ and $z=\pm h/2$. Hence the breaking strain and flexural strength are
%\begin{equation}\label{sigc-cant}
%\eps_{c}=\frac{F_{c}hL}{2YI},\ \
%\sig_{c}=Y\eps_{c}=\frac{F_{c}hL}{2I}=\frac{6F_{c}L}{bh^{2}}.
%\end{equation}

\subsection{Simple beam loaded at three points}\label{app-3pt}
%To perform an {\it in situ} cantilever test, ice is cut away from an edge to leave a rectangular beam of length $L$, width $b$ and thickness $h$, joined at one end ($x=-L/2$) to the original ice block. A downward force $F$ is then applied to the free end ($x=L/2$) until the ice breaks at the root. (When this happens $F=F_{c}$.) Then the displacement of the beam from its equlibrium position, $w$, must satisfy
Here a beam of length $L$ is simply supported  at its ends $x=\pm L/2$ and loaded in the centre ($x=0$) with a downward force $F$ until it breaks. (When this happens $F=F_{c}$.) The beam displacement $w$ satisfies
\begin{subequations}
\begin{align}
YI\pa_{x}^{4}w(x)&=-F\d(x),\\
w(\pm L/2)&=0,\label{bc-disp}\\
YI\pa_{x}^{2}w(\pm L/2)&=0.
\end{align}
\end{subequations}
Note that the delta function has units of m$^{-1}$ since its integral with respect to $x$ is 1.
Now, since the general solution to $\pa_{x}^{2}f(x)=\d(x)$ is $\displaystyle{f(x)=\frac12|x|}+Ax+B$, where $A$ and $B$ are arbitrary constants, we can immediately write $\pa_{x}^{2}w$ as
\begin{equation}\label{wxx-3pt}
\pa_{x}^{2}w(x)=\frac{F}{4YI}\big(L-2|x|  \big).
\end{equation}
(An equivalent way of formulating the equation $\pa^{2}_{x}f(x)=\d(x)$ is to require that $\pa^{2}_{x}f(x)=0$ for $|x|>0$, and that $\displaystyle{\lim_{\epsilon\ar0}\big[\pa_{x}f(x+\epsilon)-\pa_{x}f(x-\epsilon)\big]=1}$.)
We can integrate (\ref{wxx-3pt}) further and apply (\ref{bc-disp}) to find $w$, but we do not need to as we are only interested in the strain. The strain is given by
\[\eps(x)=z\pa^{2}_{x}w(x)=\frac{Fz}{4YI}\big(L-2|x|  \big),\]
which has a maximum absolute value of $\eps_{\max}=(FhL)/(8YI)$
%\begin{equation}\label{max-strain-cant}
%\eps_{\max}=\frac{FhL}{8YI}
%\end{equation}
when $x=0$ and $z=\pm h/2$. Multiplying this by $Y$ also gives the maximum stress, so the breaking strain and flexural strength are
\begin{equation}\label{sigc-3pt}
\eps_{c}=\frac h2\pa^{2}_{x}w(0)=\frac{F_{c}hL}{8YI},\ \
\sig_{c}=\frac{F_{c}hL}{8I}=\frac{3F_{c}L}{2bh^{2}}.
\end{equation}

%\subsection{Simple beam loaded at four points}\label{app-4pt}
%Here the beam is simply supported beam at its ends and loaded at two points $x=x_{1}$ and $x=x_{2}=-x_{1}$,
%equidistant from the centre ($x=0$) with downward forces of $F/2$. The  distance from each loading point to the nearest edge is $c$
%(so $x_{1}=L/2-c$). The beam displacement $w$ satisfies
%\begin{subequations}
%\begin{align}
%YI\pa_{x}^{4}w(x)&=-\frac F2\sum_{j=1}^{2}\d(x-x_{j}),\\
%w(\pm L/2)&=0,\label{bc-disp}\\
%YI\pa_{x}^{2}w(\pm L/2)&=0,
%\end{align}
%\end{subequations}
%so
%\[\pa_{x}^{2}w(x)=\frac{F}{4YI}\big(L-|x+c-L/2|-|x-c+L/2|  \big),\]
%and
%\begin{equation}\label{sigc-4pt}
%\eps_{c}=\frac h2\pa^{2}_{x}w(0)=\frac{Fhc}{4YI},\ \
%\sig_{c}=\frac{F_{c}hc}{4I}=\frac{3F_{c}c}{bh^{2}}.
%\end{equation}

\subsection{Stresses and strains in an elastic plate}\label{app-EBplate}
From \citet{fung65}, the stresses in an elastic plate with neutral plane at $z=0$ are given by

\begin{align}
\bs\sig(x,y,z,t)=-\frac{Yz}{1-\nu^{2}}
\begin{pmatrix}\Lop_{1} & \Lop_{12} & 0\\
\Lop_{12} &\Lop_{2} &  0\\
0&0& 0
\end{pmatrix}w(x,y,t),
\end{align}
where $\Lop_{1}=\nabla^{2}-(1-\nu)\pa_{y}^{2}$, $\Lop_{12}=(1-\nu)\pa_{x}\pa_{y}$, and $\Lop_{2}=\nabla^{2}-(1-\nu)\pa_{x}^{2}$.
\citet{fung65} then uses the constitutive relation
\[\bs\eps=\frac{1+\nu}Y\bs\sig-\frac{\nu}{Y}\sum_{i=1}^{3}\sig_{ii}\bfI\]
to derive the strains as
\begin{align}
\bs\eps(x,y,z,t)=-z
\begin{pmatrix}\pa_x^{2} & \pa_{x}\pa_{y} & 0\\
\pa_{x}\pa_{y} &\pa_y^{2} &  0\\
0&0& \frac{\nu}{\nu-1}\nabla^{2}
\end{pmatrix}w(x,y,t).
\end{align}

Consequently, the stresses and strains have their maximum magnitudes at the plate surfaces $z=\pm h/2$.
Also, if $w(x,y,t)$ does not vary in the $y$ direction (such as when a plane wave is traveling through the plate in a direction parallel to the $x$ axis) the maximum strain and stress (assuming $\nu<1/2$ so that $\nu/(1-\nu)<1$) are
\begin{subequations}
\begin{align}
\eps_{\max}&=\max\big\{|\eps_{11}|\big\}=\frac h2\max\big\{|\pa_{x}^{2}w|\big\},\\
\sig_{\max}&=\max\big\{|\sig_{11}|\big\}
%=\frac{Yh}{2(1-\nu^{2})}\big\{|\pa_{x}^{2}w|\big\}\notag\\&
=\frac{Y\eps_{\max}}{1-\nu^{2}},
\end{align}
\end{subequations}
taking the maximum in time over one wave period, and over the $x$ interval we are concerned with.	 Thus for $\eps_{\max}$ to exceed  $\eps_{c}=\sig_{c}/Y$ and the plate to break, we need $\sig_{\max}$ to exceed $\sig_{c}^{*}=\sig_{c}/(1-\nu)^{2}$. Consequently, we effectively have a more stringent stress condition for a plate than for a beam (where we would only require $\sig_{\max}>\sig_{c}$ for breaking to occur).

Finally, we note that for a sinusoidal traveling wave of the form (\ref{w-cos}), with wavelength $\lam_\textrm{ice}=2\pi/k_\textrm{ice}$, the maximum value the strain will ever take is
\begin{equation}\label{eps-max-app}
\eps_{\max}=\frac{h}{2}\tilde Ak_\textrm{ice}^{2}=2\pi^{2}\frac{\tilde Ah}{\lam_\textrm{ice}^{2}}.
\end{equation}

\subsection{Dispersion relation for a floating elastic plate}\label{app-disprel}
Here we present the dispersion relation for a floating elastic plate floating on water of infinite depth.
This dispersion relation appears in almost any paper dealing with wave scattering by sea ice, taking the limit as water depth goes to infinity if necessary \cite[e.g.][]{fox_squire91,squire_etal95-review}.
It assumes the water is inviscid and incompressible, that the flow is irrotational,
and that the displacement at the ice-water interface is small enough that the dynamic and kinematic boundary conditions can be {linearized}.
By coupling the resulting surface pressure with the floating elastic plate through the $p$ term in (\ref{EBplate}), this implies that a traveling wave with a surface displacement
of the form (\ref{w-cos}) is only possible if $k_\textrm{ice}$ satisfies the equation
\begin{equation}\label{disprel}
(\b_\textrm{ice} k_\textrm{ice}^{4}+\rho_\textrm{water}g-\rho_\textrm{ice}h\om^{2})k_\textrm{ice}
=\rho_\textrm{water}\om^{2}.
\end{equation}

\section{Probability density functions}\label{app-pdfs}
In the current paper we have used two main probability distributions---the Rayleigh distribution and the power-law (Pareto) distribution. \citet{kohout_meylan08} use a Rayleigh distribution to generate the random floe lengths and spacings for their ensemble averaging, but our main use for it will be in the area of wave statistics. The split (two-regime) Pareto distribution was observed by \citet{toyota-etal2011-floedist} to describe floe diameters in the MIZ around Antarctica, and \citet{dumont_etal2011} used this result to model the floe length distribution once they applied their floe breaking criteria to deduce the maximum possible floe length for a given wave field.

\subsection{Rayleigh distribution}\label{app-pdfs-ray}
The Rayleigh distribution has the PDF
%\begin{align}\label{pdf-ray}
%p_{\rm R}(D,s)=\begin{cases}
%\frac{\displaystyle{x}}{\displaystyle{\langle{D^{2}}\rangle}}\,
%\exp\left( \frac{\displaystyle{-x^{2}}}{\displaystyle{2\langle{D^{2}}\rangle}}\right),\quad&\text{for $x\geq0$}
%,\\
%0,\quad&\text{for $x<0$},
%%}
%\end{cases}
%\end{align}
\begin{align}\label{pdf-ray}
p_{\rm R}(D,s)=
\begin{cases}
\frac{\displaystyle{x}}{\displaystyle{ s^{2} }}\,
\exp\left(
\frac{\displaystyle{-x^{2}}}{\displaystyle{ 2s^{2} }}
\right),\quad&\text{for $x\geq0$}
,\\
0,\quad&\text{for $x<0$},
\end{cases}
\end{align}
%where $\langle{D^{2}}\rangle^{1/2}=\langle{D}\rangle\sqrt{2/\pi}$.
where $s=\langle{D}\rangle\sqrt{2/\pi}$.
It will be useful later to be aware that the absolute values of the extrema of a continuous normally-distributed random variable $X$ with mean zero and variance $s^{2}$ obey $p_{\rm R}(A,s)$ \citep{cartwright-longuet1956-rayleigh,longuet-higgins1980-rayleigh}. For our purposes, $A$ will be wave amplitude and the variance $s^{2}$ will be the integral of the spectral density function $S(\om)$.
%, i.e.\
%%$m_{0}$, where
%\begin{equation}
%m_{0}[w] = \int_{0}^{\infty}S(\om)\rmd\om.
%\end{equation}

\subsection{Power law distribution}\label{app-pdfs-pl}

The PDF for the truncated power law distribution is
\begin{align}\label{pdf-power-law}
p_{\rm P}(D;\g,D_{0},D_{1})=
\begin{cases}
\displaystyle{\frac{\g\b_{\rm P}}{D^{\g+1}}} \ \ &\text{if $D\in[D_{0},D_{1}]$},\\
0\ \ &\text{if $D\notin(D_{0},D_{1})$},
\end{cases}
\end{align}
where $D$ is the floe length and $\b_{\rm P}$ is chosen so that
\[\int_{D_{0}}^{D_{1}}p_{\rm P}(D)\rmd D=\b_{\rm P}\big(D_{0}^{-\g}-D_{1}^{-\g}\big)=1.\]
(The ordinary power law distribution has $D_{1}=\infty$ and $\g>0$.) This PDF gives the expected values of floe length and area (assuming that floes are approximately circular) of
\begin{subequations}
\begin{align}
\langle D\rangle&=\int_{D_{0}}^{D_{1}}p_{\rm P}(D)D\rmd D
\notag\\&
=\frac{\b_{\rm P}\g}{\g-1}\big(D_{0}^{1-\g}-D_{1}^{1-\g}\big),\label{meanD-pl}\\
\langle \mathrm{Area}\rangle
&=\frac\pi4\int_{D_{0}}^{D_{1}}p_{\rm P}(D)D^{2}\rmd D\notag\\
&=\frac{\pi\b_{\rm P}\g}{4(\g-2)}\big(D_{0}^{2-\g}-D_{1}^{2-\g}\big),\label{meanA-pl}
\end{align}
\end{subequations}
 respectively.
The split power-law distribution has the PDF
\begin{align}\label{pdf-split-power-law}
p_{\rm SP}(D;\bs\g,\bfD,\prob_{0})%\notag\\
&=\prob_{0}p_{\rm P}(D;\g_{1},D_{\rm 0},D_{1})\notag\\
&\quad + (1-\prob_{0})p_{\rm P}(D;\g_{2},D_{ 1},\infty).
\end{align}
Apart from the exponents in $\bs\g=(\g_{1},\g_{2})$ and the diameters in $\bfD=(D_{0},D_{1})$, there is an additional free parameter in this PDF, $\prob_{0}=\prob(D<D_{1})$.

\section*{Acknowledgments}
This work is part of the project `Waves-in-Ice Forecasting for Arctic operators' (WIFAR), funded by the Norwegian Foundation of Research and Total E\&P. We wish to thank Aleksey Marchenko, Gareth Hegarty and David Leslie for helpful discussions, and Francois Counillon for assistance with the TOPAZ data and related discussions.

\end{document}